\documentclass[preprint2]{aastex}

\usepackage[dvips]{color}

\slugcomment{}

\shorttitle{Protogalaxy Evolution}
\shortauthors{Yamasawa et al.}

\begin{document}

\title{The Role of Dust in the Early Universe I : \\
Protogalaxy Evolution}
\author{Daisuke Yamasawa\altaffilmark{1}, Asao Habe and Takashi Kozasa}
\affil{Department of Cosmosciences, Graduate School of Science, Hokkaido University, Sapporo, Hokkaido 060-0810, Japan}
\author{Takaya Nozawa}
\affil{Institute for the Physics and Mathematics of the Universe, University of Tokyo, Kashiwa, Chiba 277-8583, Japan}
\author{Hiroyuki Hirashita}
\affil{Institute of Astronomy and Astrophysics, Academia Sinica, P.O. Box 23-141, Taipei 10617, Taiwan}
\author{Hideyuki Umeda}
\affil{Department of Astronomy, School of Science, University of Tokyo, Hongo, Tokyo 113-0033, Japan}
\and
\author{Ken'ichi Nomoto}
\affil{Institute for the Physics and Mathematics of the Universe, University of Tokyo, Kashiwa, Chiba 277-8583, Japan}

\altaffiltext{1}{yamasawa@astro1.sci.hokudai.ac.jp}
\begin{abstract}
We develop one-zone galaxy formation models in the early Universe, taking into account dust formation and evolution by supernova (SN) explosions. 
Especially we focus on the time evolution of dust size distribution, because ${\rm H}_{2}$ formation on the dust surface plays a critical role in the star formation process in the early Universe. 
In the model we assume that star formation rate (SFR) is proportional to the total amount of ${\rm H}_{2}$. 
We consistently treat (i) the formation and size evolution of dust, (ii) the chemical reaction networks including ${\rm H}_{2}$ formation both on the surface of dust and in gas phase, and (iii) the SFR in the model. 
First, we find that, because of dust destruction due to both reverse and forward shocks driven by SNe, H$_{2}$ formation is more suppressed than that without dust destruction. 
At the galaxy age of $\sim0.8\ {\rm Gyr}$, for galaxy models with virial mass $M_{\rm vir}=10^{9}\ M_{\odot}$ and formation redshift $z_{\rm vir}=10$, the molecular fraction is 2.5 orders of magnitude less in the model with dust destruction by both shocks than that in the model without dust destruction. 
Second, we show that the H$_{2}$ formation rate strongly depends on the ISM density around SN progenitors. 
The SFR in higher ISM density is lower, since dust destruction by reverse shocks is more effective in higher ISM density. 
We conclude that not only the amount but also the size distribution of dust being related with the star formation activity strongly affects the evolution of galaxies in the early Universe. 
\end{abstract}





\section{Introduction}
Understanding of galaxy evolution in the early Universe remains one of the most important goals of modern astrophysics. 
Modeling of primeval galaxy formation requires an accurate treatment of star formation process in low-metallicity gas \citep{Jap07, Glo07, Smi08, Smi09, Jap09a, Jap09b}. 
A critical challenge for achieving this goal is due to our poor understanding of how gas is converted into stars under different conditions \citep{Kru05, Rob08, Gne09}. 
In particular, star formation efficiency in primeval galaxy is still uncertain. 

The standard approach in theoretical studies of galaxy formation so far is to adopt a recipe which ties the star formation rate (SFR) to gas density both in semi-analytic models \citep[e.g.][]{Col00} and in numerical simulations \citep[e.g.][]{Spr05}. 
Such a recipe is based on the empirical correlations observed in local galaxies, namely the Kennicut-Schmidt law \citep{Ken98}. 
These correlations have only been studied relatively well for nearby massive or star bursting galaxies. 
However, for galaxies with low surface brightness and/or low-metallicity, this empirical relation may not be valid. 
Indeed, both nearby metal-poor galaxies \citep{Big08} and high-redshift galaxies \citep{Wol06} provide a variety of clues suggesting that gas conversion into stars in low-mass, low-metallicity galaxies is very inefficient. 

The star formation efficiency may depend on ability to convert a fraction of gas mass into molecular form. 
Molecular hydrogen is produced by chemical reactions in gas phase in first galaxy halos. 
In the reionization era, ${\rm H}_{2}$ molecule dissociation by the Lyman-Werner ultraviolet (UV) background between 11.2 and 13.6 eV is important in the lower mass ${\rm H}_{2}$ cooling halos. 
Gas condensation in the lower mass ${\rm H}_{2}$ cooling halos can be delayed by the Lyman-Werner background \citep{Mac01, Mac03, Yos03, Sus07, Wis07, O'S08}. 
The Lyman-Werner background thus increases cooling times in the centers of such halos. 
As a result, the minimum mass of a star-forming halo increases with the Lyman-Werner background intensity. 
The Lyman-Werner background becomes less of an issue in atomic line cooling halos as Ly$\alpha$ cooling provides ample amounts of free electrons for ${\rm H}_{2}$ cooling, and they become self-shielding to this radiation \citep{O'S08, Sus08, Wis08, Wis09}. 

In the later epoch, dust ejected by stars in galaxies is effective to shield the Lyman-Werner background and acts as an effective catalyst for ${\rm H}_{2}$ molecule production on the dust grains. 
In simulations with star formation models based on molecular hydrogen \citep{Rob08, Gne09}, once the gas enriched up to $Z\sim0.01-0.1\ {Z_{\odot}}$, the subsequent star formation and enrichment of metal and dust can be much more accelerated. 
\citet{Gne09} show that the transition from atomic to molecular hydrogen depends primarily on metallicity, assuming that the dust abundance is directly related to metallicity. 

Dust plays a crucial role in the star formation: 
(i) molecular hydrogen is produced more efficiently on dust grains than in gas phase, 
(ii) dust shields dissociating UV radiation, 
and (iii) dust allows the formation of low-mass stars in low-metallicity environments, and hence affects the initial mass function (IMF) \citep{Omu05, Sch06, Sch10, Omu10}. 

In theoretical studies on the molecular abundance in the interstellar medium (ISM), dust abundance is often scaled with the metallicity and dust grain properties are assumed to be the same as in the local ISM. 
However, the composition of dust is likely to be different in early galaxies. 
The observational evidence is that the dust extinction curves of the broad absorption line quasars at $z>4$ are likely to be due to the type II SN (SN II) dust \citep{Mai04, Gal10}. 

Since the lifetime of SN II progenitor is short, SN II can be the dominant production source of dust grains in young ($<1$ Gyr) galaxies. 
Primeval SNe produced by Population III stars \citep{Bro03, Kit05, Wha08} may contribute the dust production \citep{Noz03, Sch04}. 
The winds of evolved low-mass stars contribute to dust formation considerably in nearby galaxies, but the cosmic time is not long enough for such stars to evolve at high redshift ($z>5$) where all galaxies should have ages younger than $\simeq 1$ Gyr. 
Contribution of dust production by low-mass stars is not dominant in such young galaxies. 
In addition, dust is destroyed by SN shocks. 
Thus, the modeling of dust evolution in galaxies requires an accurate treatment of production and destruction of dust grains together with star formation activities \citep{Hir02}. 

In this paper, we investigate not only the evolution of dust mass but also the time evolution of dust size distribution. 
The dust size distribution evolves rapidly because of the destruction by sputtering in the high-velocity shocks driven by SNe. 
Collision of the expanding SN ejecta with the surrounding ISM creates a forward shock at the interface between the ejecta and the ISM \citep{Noz06}, and a reverse shock that penetrates into the ejecta \citep{Bia07, Noz07, Nat08, Sil10}. Since the erosion rate by sputtering does not strongly depend on the grain size, small grains are predominantly destroyed regardless of grain species. 
Therefore, the fraction of small size grains relatively decreases with galaxy evolution. 

We focus on the effects of molecular hydrogen abundance on the SFR in the early stage of galaxy evolution, taking into account molecular formation on dust, since ${\rm H}_{2}$ formation on dust surface is very effective \citep{Hir02, Caz04}. 
\citet{Hir02} show that this effect causes an enhancement of the SFR by an order of magnitude on a timescale of $3-5$ galactic dynamical time. 
However, they assumed a single dust grain size ($\sim0.03\ \mu{\rm m}$). 
We adopt more accurate analytic formulae for the formation of molecular hydrogen on dust grains than \citet{Hir02} by using the results of dust size distribution by \citet{Noz06, Noz07}. 

This is the first study on galaxy evolution considering dust size evolution for halo masses above $10^{8-9}$ in the high-redshift ($5<z<10$), whose interiors we expect to be roughly self-shielded from both ionizing and Lyman-Werner UV radiation. 
To show clearly the dependence of galaxy properties on dust destruction, we use a simple one-zone galaxy model. 
%and study the case of atomic cooling halos. 

The paper is organized as follows. 
In $\S2$ we describe the dust evolution model. 
In $\S3$ we explain our one-zone galaxy model. 
In $\S4$ we present the results.
In $\S5$ we discuss the effects of the dust size evolution on H$_{2}$ formation process and conclude by summarizing our results. 
Throughout this paper we adopt the cosmological parameters from the third-year $WMAP$ results \citep{Spe07}, $\Omega_{\Lambda}=0.76$, $\Omega_{M}=0.24$, $\Omega_{b}=0.04$, and $H_{0}=73\ {\rm km}\  {\rm s}^{-1}\ {\rm Mpc}^{-1}$. 

\section{Dust evolution model}
\label{sec:dustmodel}
\subsection{Source of dust in the early Universe}
\label{subsec:sorceofdust}
SNe II are believed to be the dominant sources of dust at high redshift of $z>5$ because of short lifetimes ($<10^{7}\ {\rm yr}$) of their massive progenitors \citep[e.g.][]{Dwe07, Gal10}. 
Dust formation in the ejecta of primordial SNe II has been investigated theoretically \citep{Tod01, Noz03, Che10}.
The amount and the size distributions of dust grains injected into ISM have been investigated by considering the destruction in SN remnants (SNRs) \citep{Bia07, Noz07, Nat08, Sil10}. 
Although how much dust really forms in the ejecta has been still under debate \citep[][for review]{Koz09}, the recent observations of Cas-A SNR revealed the presence of $\sim0.07\ M_{\odot}$ dust condensed in the ejecta \citep{Bar10, Sib10}, which is consistent with the dust mass predicted by the theoretical model taking into account formation and destruction processes of dust in a Type IIb SN \citep{Noz10}. 
\citet{Val09} and \citet{Dwe10} have proposed that the contribution from asymptotic giant branch (AGB) stars in the high-redshift quasar J1148+5251 cannot be neglected for the total dust budget even at $z\sim6$. 
However, the size distribution of dust formed in the mass-loss wind of AGB stars has not been fully studied yet \citep{Fer06, Zhu08}. 
If type Ia SNe could occur in such an early epoch, they are unlikely to be efficient sources of dust (Nozawa et al. 2010 in preparation). 
Therefore, in order to follow the evolution of dust size distribution and reveal the resulting influence on galaxy evolution, we consider SNe II as the source of dust in the early Universe. 

The basic quantity for governing the production and destruction history of dust by SNe II is the rate of SN II explosions, $\gamma_{\rm SN}(t)$, given by
\begin{equation}
	\gamma_{\rm SN}(t)=\int_{m_{\rm SN}^{l}}^{m_{\rm SN}^{u}}{\rm d}m\Psi(t-\tau(m))\phi(m),
\label{eq:gam}
\end{equation}
where $\Psi(t)$ is the SFR at time $t$, $\phi(m)$ is the stellar IMF, $\tau(m)$ is the lifetime of a star whose mass is $m$, and $m_{\rm SN}^{u}$ and $m_{\rm SN}^{l}$ are the upper and lower mass limits of SN II progenitors, respectively. 
In this paper we adopt the Salpeter IMF ($\phi(m)\propto m^{-2.35}$, \citep{Sal55}) with the stellar mass range between $0.1\ M_{\odot}$ and $60\ M_{\odot}$, we assume $m_{\rm SN}^{l}=8\ M_{\odot}$ and $m_{\rm SN}^{u}=40\ M_{\odot}$ \citep{Her03}. 
For $\tau(m)$, we adopt the model of zero-metallicity stars without mass loss \citep{Sch02}. 

In this paper, we do not consider Population III stars, for simplicity. 
In the forthcoming paper, we will consider possible contribution of Population III stars. 
Population III stars formed out of the primordial gas are considered to be much more massive than Population I/II stars \citep[][for reviews]{Yos08, Bro09}, and thus the primordial IMF might be biased toward a higher mass ($\gtrsim10\ M_{\odot}$) than that in the present Universe. 
Furthermore, Population III stars as massive as $140-260\ M_{\odot}$ are predicted to end their lives as pair-instability SNe \citep[PISNe][]{Her02} and to produce a large amount of metals and dust \citep{Noz03, Sch04}. 
However, \citet{Jog10a} and \citet{Jog10b} address the growing nucleosynthetic 'forensic' evidence that the majority of primordial stars may have been $15-40\ M_{\odot}$ objects. 
On the other hand, once the gas is enriched up to a critical metallicity of $Z\simeq10^{-6}-10^{-5}\ Z_{\odot}$, formation of low-mass stars is triggered, leading to the transition of the star formation mode from massive population III stars to low-mass population I/II stars, if dust present \citep{Omu05, Sch06, Sch10, Omu10}. 
If there is no dust, the transition of the star formation mode is expected to occurs at $10^{-3.5}\ Z_{\odot}$ \citep{Mack03}. 
In this case, the CMB limits the lower masses of stars to a few $10$s of $M_{\odot}$  \citep{Smi08, Sch10}. 
Formation history of galaxies considering the time-dependent IMF from the top-heavy to the Salpeter-like IMF and taking into account the production and destruction of dust by PISNe, will be explored in the forthcoming paper (Yamasawa et al. 2010 in preparation). 

\subsection{Dust injected from SNe II into ISM}
\label{subsec:dustprod}
Throughout this paper we adopt the models by \citet{Noz03, Noz07} for dust formation and destruction. 
\citet{Noz03} investigated the dust production in the ejecta of primordial SNe II as well as PISNe, applying a theory of non-steady state nucleation and grain growth. 
They revealed the grain species formed in the ejecta and their size distributions for the unmixed and mixed elemental compositions within the He core. 
In what follows, we apply the results of calculation for the unmixed ejecta of SNe II with the progenitor mass $m=13, 20, 25,\ {\rm and}\ 30\ M_{\odot}$ and the explosion energy $10^{51}\ {\rm erg}$, and extrapolate the data to the mass range from $8$ to $40\ {M_{\odot}}$. 
\begin{figure}[h]
\epsscale{1.0}
\plotone{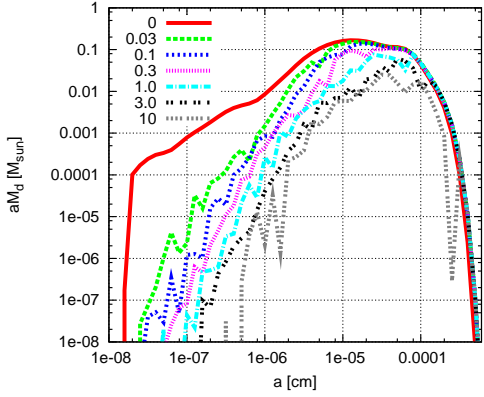}
\caption{The IMF-averaged mass distributions of dust formed in the ejecta and injected into ISM per SN II:
The solid line denotes the mass distribution of dust before the destruction through a reverse shock, $a\overline{{\cal M}_{\rm d}^{0}}(a)$ and the dashed lines and the dotted lines denote the mass distributions of dust at the injection into ISM after the destruction through a reverse shock, $a\overline{{\cal M}_{\rm d}^{n_{\rm SN}}}(a)$, for the number densities of gas around a SN progenitor, $n_{\rm SN}=0.03$, $0.1$, $0.3$, $1.0$, $3.0$, and $10.0$ ${\rm cm}^{-3}$, which are used to annotate the curves. 
The horizontal axis shows the radius of dust in units of cm. 
The vertical axis is the mass distribution of dust $a\overline{{\cal M}_{\rm d}^{n_{\rm SN}}}(a)$ in units of $M_{\odot}$.} 
\label{fig:prod}
\end{figure}

The solid line in Figure \ref{fig:prod} shows the IMF-averaged mass distribution of dust formed in the ejecta of SNe II, $\overline{{\cal M}_{{\rm d}}^{0}}(a)$, which is weighted by the Salpeter IMF and is summed up over all the grain species as, 
\begin{eqnarray}
	\overline{{\cal M}_{\rm d}^{0}}(a)&=&\sum_{j}\overline{{\cal M}_{{\rm d},j}^{0}}(a) \nonumber\\
	&=&\frac{\sum_{j}\int_{{\rm m}_{\rm SN}^{l}}^{m_{\rm SN}^{u}}{\rm d}m\ {\cal M}_{{\rm d},j}^{0}(a,m)\phi(m)}{\sum_{j}\int_{m_{\rm SN}^{l}}^{m_{\rm SN}^{u}}{\rm d}m\ \phi(m)},
	\label{proddustnoreverse}
\end{eqnarray}
where ${\cal M}_{{\rm d},j}^{0}(a,m){\rm d}a$ is the mass of the $j$-th dust species produced in a SN II with radii between $a$ and $a+{\rm d}a$ as a function of progenitor mass $m$, and superscript 0 means the case with no destruction by a reverse shock. 
In Figure \ref{fig:prod}, we plot $a\overline{{\cal M}_{{\rm d}}^{0}}(a)$ in the vertical axis to make clear the mass fraction in each logarithmic bin. 
We can see that the grain radii range from a few \AA\ up to a few $\mu{\rm m}$ and that the size spectrum of dust in mass has a peak at $a\sim0.1\ \mu{\rm m}$. 

In the course of their injection into ISM, dust grains formed in the ejecta are destroyed due to sputtering in the hot gas between the reverse and forward shocks, which is hereafter referred to as the destruction by reverse shock. 
\citet{Noz07} investigated the survival of the newly formed dust in the shocked gas within the SNRs expanding into the uniform ISM with hydrogen number densities of $n_{\rm SN}=0.1$, $1.0$ and $10.0\ {\rm cm^{-3}}$, and showed that the destruction efficiency of newly formed dust is not only sensitive to the initial size distribution but also strongly depends on $n_{\rm SN}$. 
To investigate the dependence of destruction of dust on the ISM densities, we extend their models to six cases of $n_{\rm SN}=0.03$, $0.1$, $0.3$, $1.0$, $3.0$, and $10.0$ ${\rm cm}^{-3}$. 

Figure \ref{fig:prod} shows the IMF-averaged mass distribution of dust, $\overline{{\cal M}_{{\rm d}}^{n_{\rm SN}}}(a)$, injected into ISM after destruction by the reverse shock for $n_{\rm SN}=0.03$, $0.1$, $0.3$, $1.0$, $3.0$, and $10.0$ ${\rm cm}^{-3}$, which is weighted by the Salpeter IMF and is summed up over all the grain species as in Equation (\ref{proddustnoreverse}). 
The mass distribution of the $j$-th dust species for the case of $n_{\rm SN}$, $\overline{{\cal M}_{{\rm d},j}^{n_{\rm SN}}}(a)$, is the IMF-averaged one after the destruction by the reverse shock. 
We can see that the change in the dust mass distribution through processing in SNRs becomes more (less) prominent for higher (lower) gas density; small size grains get deficient with increasing $n_{\rm SN}$. 
The dust grains with radii below $0.01\ {\rm \mu m}$ are preferentially destroyed by sputtering for $n_{\rm SN}>0.03\ {\rm cm}^{-3}$, while dust with radii larger than $\sim1\ {\rm \mu m}$ are almost intact for $n_{\rm SN}\le3.0\ {\rm cm}^{-3}$. 
As a result, the mass of dust injected into the ISM is fully dominated by grains with radii above $\sim0.1\ {\rm \mu m}$. 

The total geometrical cross-section of dust per metal mass is an important quantity for ${\rm H}_{2}$ formation on grain surface at a certain metallicity level. 
The total geometrical cross-section of dust injected into the ISM from a SN depends on the gas density around the SN progenitor, $n_{\rm SN}$. 
In many papers, the total geometrical cross-section of dust is scaled to metal mass under following two assumptions: (i) the depletion factor which is defined as the dust mass per metal mass, is identical to that in the MW, and (ii) the dust size distribution which determines dust area per unit dust volume $\langle a^{2}\rangle/\langle a^{3}\rangle$, is the same as that in the those in the MW. 
However, the depletion factor and the size distribution of the SN II dust are quite different from those in the MW. 
\begin{figure}[!h]
\epsscale{1.0}
\plotone{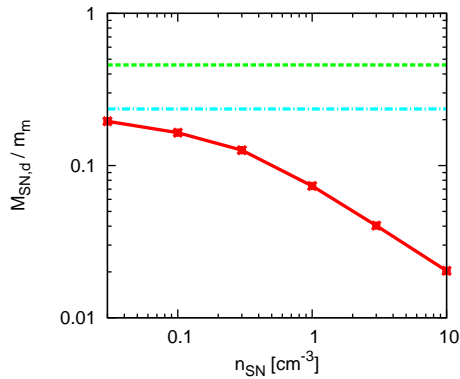}
\caption{The depletion factor, $\overline{M_{\rm SN,d}^{n_{\rm SN}}}/\overline{m_{m}}$, for various dust destruction models with the densities of ISM around the SN II progenitor, $n_{\rm SN}=0.03, 0.1, 0.3, 1, 3, 10\ {\rm cm}^{-3}$. 
The dot-dashed line represents the depletion factor, $\overline{M_{\rm SN,d}^{0}}/\overline{m_{m}}$ for the model without reverse shock destruction. 
The ejected metal mass per SN II is taken from \citet{Ume02}. 
The dotted line represents the typical ratio in MW, where we assume a dust-to-gas mass ratio to be ${\cal D}=0.00934$ \citep{Pol94}, and assume a ratio of metal mass to hydrogen mass to be $0.0204$ \citep{Omu00}. }
\label{dustdepl}
\end{figure}

Figure \ref{dustdepl} shows the depletion factor, $\overline{M_{\rm SN,d}^{n_{\rm SN}}}/\overline{m_{m}}$ after the destruction by reverse shock as a function of ISM density $n_{\rm SN}$, where $M_{\rm SN,d}^{n_{\rm SN}}$ is the IMF-averaged total dust mass ejected into ISM by a SN, 
\begin{equation}
	\overline{M_{\rm SN,d}^{n_{\rm SN}}}=\int_{0}^{\infty}{\rm d}a\overline{{\cal M}_{\rm d}^{n_{\rm SN}}}(a),
\end{equation}
and $\overline{m_{m}}$ is the IMF-averaged total metal mass ejected into ISM, 
\begin{eqnarray}
	\overline{m_{m}}&=&\sum_{i}\overline{m_{{\rm m},i}} \nonumber\\
	&=&\frac{\sum_{i}\int_{m_{\rm SN}^{l}}^{m_{\rm SN}^{u}}m_{{\rm m},i}(m)\phi(m){\rm d}m}{\sum_{i}\int_{m_{\rm SN}^{l}}^{m_{\rm SN}^{u}}\phi(m){\rm d}m}. 
\end{eqnarray}
$m_{{\rm m},i}(m)$ is the mass of $i$-th element of metal ejected from SN with progenitor mass is $m$ and is taken from \citet{Ume02}. 
The depletion factor in the case of $n_{\rm SN}=1.0\ {\rm cm}^{-3}$ is a factor 6 smaller than that in the MW. 
\begin{figure}[!h]
\epsscale{1.0}
\plotone{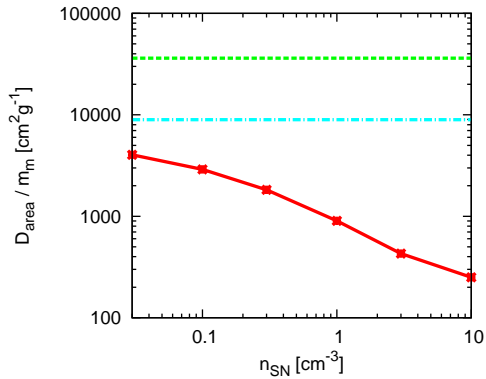}
\caption{The ratio of total dust cross-section to total metal mass, $\overline{D_{\rm area}^{n_{\rm SN}}}/\overline{m_{\rm m}}$, for various dust destruction models with the densities of ISM around the SN II progenitor, $n_{\rm SN}=0.03, 0.1, 0.3, 1, 3, 10\ {\rm cm}^{-3}$. 
The dot-dashed line represents the ratio, $\overline{D_{\rm area}^{0}}/\overline{m_{\rm m}}$ for the model without reverse shock destruction. 
The ejected metal mass per SN II is taken from \citet{Ume02}. 
The dotted line represents the ratio in our galaxy, where we assume a dust properties as a size distribution, $f(a)\propto a^{-3.5}$ $(5\times10^{-7}\ {\rm cm}<a<2.5\times10^{-5}\ {\rm cm})$ \citep{Mat77, Dra84} and a dust-to-gas mass ratio, ${\cal D}=0.00934$ \citep{Pol94}, and assume that a ratio of metal mass to hydrogen mass is $0.0204$ and the bulk density of dust is $3.0\ {\rm g\ cm}^{-3}$ \citep{Omu00}. }
\label{dustareatometal}
\end{figure}

Figure \ref{dustareatometal} shows the ratio of total geometric cross-section of dust to the total metal mass, $\overline{D_{\rm area}^{n_{\rm SN}}}/\overline{m_{\rm m}}$ after the destruction by the reverse shock as a function of ISM density $n_{\rm SN}$. 
The total geometrical cross-section of dust weighted by the Salpeter IMF, $\overline{D_{\rm area}^{n_{\rm SN}}}$, is written as
\begin{equation}
	\overline{D_{\rm area}^{n_{\rm SN}}}=\sum_{j}\frac{3}{4\rho_{j}}\int_{0}^{\infty}{\rm d}a\overline{{\cal M}_{{\rm d},j}^{n_{\rm SN}}}(a)/a,
\end{equation}
where $\rho_{j}$ is the bulk density of $j$-th dust species. 
The ratio, $\overline{D_{\rm area}^{n_{\rm SN}}}/\overline{m_{\rm m}}$, is smaller for larger $n_{\rm SN}$ because small dust grains are efficiently destroyed by sputtering in the SNR under large $n_{\rm SN}$; note that the surface area per dust mass is larger for smaller-size grains. 
Also with increasing $n_{\rm SN}$, the reverse shock becomes stronger and destroys dust by sputtering more effectively \citep[see][]{Noz07}. 
In addition we plot the ratio for the case without reverse shock together with the typical value in the MW for comparison. 
$\overline{D_{\rm area}^{n_{\rm SN}}}/\overline{m_{\rm m}}$ in the model without reverse shock is a factor of 4 smaller than that in the MW. 
$\overline{D_{\rm area}^{n_{\rm SN}}}/\overline{m_{\rm m}}$ in the case with $n_{\rm SN}=1.0\ {\rm cm^{-3}}$ is 40 times smaller than that in the MW. 
Thus, the rescaling of cross-section of dust by the metal mass using the MW value results in significant overestimate for ${\rm H}_{2}$ formation in high-redshift galaxies. 

\subsection{Destruction of interstellar dust by SN forward shocks}
\label{subsec:dustdest}
Dust grains injected into the ISM are subjected to destruction by the blast waves (the high-velocity interstellar shocks) driven by the ambient SNe \citep[e.g.][]{Jon94}. 
\citet{Noz06} investigated the processing of interstellar dust by sputtering in the hot gas swept up by the SN forward shock. 
Adopting the dust model by \citet{Noz03} as the size distribution of interstellar dust, \citet{Noz06} have shown that the destruction efficiency of dust depends on the ISM density and the explosion energy of SNe as well as the initial size distribution of dust. 
It should be noted that the size distribution as well as the destruction efficiency changes as a function of time because interstellar dust are supplied from SNe and processed in ISM successively according to star formation activity. 
Thus, we must deal with the destruction process in a way that is applicable to any dust size distribution to explore the global evolution of dust size distribution. 

In order to evaluate the destruction efficiency of interstellar dust for any initial size distribution, here we introduce the conversion efficiency as defined below. 
Consider that the $j$-th dust species residing in the ISM, whose size distribution is given by the number of dust grains with radii between $a$ and $a+{\rm d}a$, $f_{j}(a){\rm d}a$, is processed by sputtering in hot plasma produced through the a single passage of SN shock. 
The conversion efficiency $\eta_{j}(a,a^{\prime})$ is defined as the number fraction of dust grains with radii between $a^{\prime}$ and $a^{\prime}+{\rm d}a^{\prime}$ that are converted to grains with radii between $a$ and $a+{\rm d}a$ by sputtering through the passage of a SN shock. 
The number of dust grains with radii between $a$ and $a+{\rm d}a$  produced by the sputtering is given as $\eta_{j}(a,a^{\prime})f_{j}(a^{\prime}){\rm d}a^{\prime}$. 
Note that $\eta_{j}(a,a^{\prime})=0$ for $a>a^{\prime}$. 
Then the change in the number of dust grains with radii between $a$ and $a+{\rm d}a$ caused by a shock processing is given by
\begin{eqnarray}
	{\rm d}N_{j}(a)&=&\sum_{a^{\prime}>a}^{\infty}\eta_{j}(a,a^{\prime})f_{j}(a^{\prime}){\rm d}a^{\prime} \nonumber\\
	&&-\left[1-\eta_{j}(a,a)\right]f_{j}(a){\rm d}a \nonumber\\
	&=&\int_{0}^{\infty}\eta_{j}(a,a^{\prime})f_{j}(a^{\prime}){\rm d}a^{\prime}-f_{j}(a){\rm d}a, \nonumber\\
\end{eqnarray}
and as well the corresponding change of the mass is given by
\begin{eqnarray}
	{\rm d}M_{{\rm d},j}(a)&=&\frac{4\pi}{3}a^{3}\rho_{j}\int_{0}^{\infty}\eta_{j}(a,a^{\prime})f_{j}(a^{\prime}){\rm d}a^{\prime} \nonumber\\
	&&-{\cal M}_{{\rm d},j}(a){\rm d}a
\end{eqnarray}
where ${\cal M}_{{\rm d},j}(a){\rm d}a$ is the mass of the pre-shocked dust. 
The size distribution function after the shock processing $f^{\prime}_{j}(a)$ is given by $f^{\prime}_{j}(a)=f_{j}(a)+{\rm d}N_{j}/{\rm d}a$. 

The conversion efficiency $\eta(a,a^{\prime})$ and the mass of ISM gas swept up by shock $M_{\rm swept}$ depend on the progenitor mass, expanding energy and type of SN as well as the structure, number density and metallicity of ISM gas. 
For these parameters of SNe and ambient ISM, once $M_{\rm swept}$ and $\eta(a,a^{\prime})$ for each dust species are calculated, the time evolution of dust mass and size distribution can be followed in a consistent way with the star formation activity in galaxies as described in Section \ref{sec:galmodel}. 

The calculations of $\eta(a,a^{\prime})$ and $M_{\rm swept}$ are performed by using the method developed by \citet{Noz06} as follows; 
the efficiency of dust destruction increases with increasing the explosion energy and/or increasing $n_{\rm SN}$ but is almost independent of the progenitor mass as long as the explosion energy is the same \citep{Noz06}. 
We assume that SNe driving high-velocity shock in ISM are represented by Type II SN with the progenitor mass of $20\ M_{\odot}$ and the explosion energy of $10^{51}\ {\rm erg}$. 
The ISM surrounding the SN is considered to be uniform with hydrogen number densities $n_{\rm SN}=0.03$, $0.1$, $0.3$, $1.0$, $3.0$ and $10\ {\rm cm}^{-3}$. 
By distributing dust grains with radius $a^{\prime}$ uniformly in ISM, the conversion efficiency $\eta(a,a^{\prime})$ is evaluated for each grain species by calculating the erosion of dust by sputtering until the truncation time $t_{\rm tr}$ which is defined as a time when the shock velocity is decelerated below $100\ {\rm km}\ {\rm s}^{-1}$ \citep[see][for the details]{Noz06}. 
In the calculations, the radii of grains in the ISM range from $0.00013$ to $6.3\ {\rm \mu m}$ for each grain species. 
\begin{figure}[h]
\epsscale{1.0}
\plotone{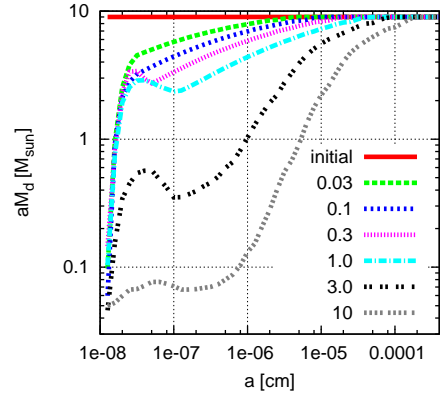}
\caption{The dust mass, $\sum_{j}a{\cal M}_{{\rm d},j}(a)$, processed by a forward shock driven by a SN II explosion in the ambient ISM gas density of $n_{\rm SN}=0.03$, $0.1$, $0.3$, $1.0$, $3.0$, and $10.0\ {\rm cm}^{-3}$,which are used to annotate the curves, and metallicity of $Z=10^{-4}\ Z_{\odot}$. 
We assume the initial size distribution to be $a{\cal M}_{{\rm d},j}(a)=1$ for each species $j$. 
In the calculations, we use the conversion efficiency $\eta_{j}(a,a^{\prime})$ evaluated for SN explosion energy, $10^{51}\ {\rm erg}$, and progenitor mass, $m=20\ M_{\odot}$. }
\label{fig:dest}
\end{figure}

In Figure \ref{fig:dest}, we present the changes in the dust size distributions due to the shock-processing for different ISM-densities, where the initial mass distribution of dust is set to be $a{\cal M}_{{\rm d},j}(a)=1$ for clarity. 
As can be seen from the figure, small-size grains are destroyed significantly due to the erosion by sputtering, and more dust grains are processed for a higher ISM density. 

The mass of gas swept up by the forward shock until the truncation time $t_{\rm tr}$, $M_{\rm swept}$, depends on not only the ISM density but also the initial metallicity of the gas in the ISM. 
As the line cooling by heavy elements becomes more efficient for a higher gas metallicity, the forward shock is decelerated more quickly, resulting in a smaller $M_{\rm swept}$. 
By fitting $M_{\rm swept}$ calculated for different $n_{\rm SN}$ and $Z$, we derived the following approximation formula, 
\begin{equation}
M_{\rm swept}/M_{\odot}=1535n_{\rm SN}^{-0.202}\left[\left(Z/Z_{\odot}\right)+0.039\right]^{-0.298},
\end{equation}
whose fitting accuracy is within $16\%$ for $0.03\ {\rm cm}^{-3}\le n_{\rm SN}\le 30\ {\rm cm}^{-3}$ and for $10^{-4}\le Z/Z_{\odot}\le1.0$. 

\subsection{Formulation of dust size evolution}
In terms of the conversion efficiency describing the processing of dust by sputtering, here we formulate the time evolution of the mass of $j$-th dust grains with radii between $a$ and $a+\Delta a$ in our model galaxies, $\Delta M_{{\rm d},j}(a,t)=\frac{4\pi}{3}a^{3}\rho_{j}f_{j}(a,t)\Delta a$, as 
\begin{eqnarray}
	\frac{{\rm d}\Delta M_{{\rm d},j}(a,t)}{{\rm d}t}&=&\overline{\Delta M_{{\rm SN,d},j}}(a)\gamma_{\rm SN}(t) \nonumber\\
	&&-\frac{M_{\rm swept}}{M_{\rm ISM}(t)}\gamma_{\rm SN}(t)\times\{\Delta M_{{\rm d},j}(a,t) \nonumber\\
	&&-\int_{0}^{\infty}{\rm d}a^{\prime}\eta_{j}(a,a^{\prime})f_{j}(a^{\prime},t)\rho_{j}\frac{4\pi}{3}a^{3}\} \nonumber\\
	&&-\Psi(t)\frac{\Delta M_{{\rm d},j}(a,t)}{M_{\rm ISM}(t)}
	\label{dustformulation}
\end{eqnarray}
where $M_{\rm ISM}(t)$ is the total mass of gas and dust, and $f_{j}(a,t)$ is the size distribution function of dust species $j$ in the ISM at a time $t$. 
We note that $(M_{\rm swept}\gamma_{\rm SN}(t)/M_{\rm ISM}(t))^{-1}$ is a timescale of sweeping whole ISM by SNe. 
The IMF-averaged mass of dust species $j$ with radii between $a$ and $a+\Delta a$ injected from SNe II into the ISM with number density $n_{\rm ISM, SN}$ is defined by $\overline{\Delta M_{\rm SN,d,j}}(a)=\overline{{\cal M}_{\rm d,j}^{n_{\rm SN}}}(a)\Delta a$. 
The first term on the right-hand side is the injection rate of dust from SNe II. 
The second term is the destruction rate of interstellar dust by SN blast waves, and the third term is the rate at which the interstellar dust is incorporated into stars. 

\section{Galaxy evolution model}
\label{sec:galmodel}
\subsection{Dark matter halo and physical state of gas}
\label{subsec:formdm}
We quantify the properties of dark matter halos, assuming a dynamically equilibrium state.
The radius of dark matter halo, $r_{\rm vir}$, is estimated in terms of the mass of dark halo, $M_{\rm vir}$, and the redshift of virialization, $z_{\rm vir}$, as 
\begin{equation}
	\frac{4}{3}\pi r_{\rm  vir}^{3}\left\{1+\delta_{\rm  c}(z_{\rm vir})\right\}\rho_{{\rm  c}0}\Omega_{M}(1+z_{\rm  vir})^{3}=M_{\rm  vir},
\end{equation}
where $\rho_{{\rm c}0}\equiv3H_{0}^{2}/8\pi G$ is the critical density of the Universe at $z=0$, $\delta_{c}(z_{\rm vir})$ is the overdensity of a dark matter halo vilialized at $z_{\rm vir}$, and $G$ is the gravitational constant. 

We assume dark matter halos as singular isothermal spheres and rotating uniform gas disks in their gravitational potentials. 
Cosmological N-body simulations show that structures of dark matter halos are well described by the NFW profile \citep{Nav96}. 
\citet{Mo98} studied a simple disk model in the gravitational potential of the singular isothermal sphere and more realistic disk model in the gravitational potential of the NFW halo profile. 
We adopt a radius of the disk, $r_{\rm disk}\simeq0.18r_{\rm vir}$ \citep{Fer00,Hir02}, by considering the conservation of angular momentum and assuming a typical value for the spin parameter $\lambda=0.04$ from the paper by \citet{Fer00} who estimate the radius of the disk as $r_{\rm disk}=4.5 \lambda r_{\rm vir}$ in a modified isothermal halo. 

In our one-zone model, we need a virial temperature for the initial gas temperature and a dynamical timescale of gas in the disk. 
Gas collapsed at $z_{\rm vir}$ in the dark matter halo of $M_{\rm vir}$ has a virial temperature, $T_{\rm vir}$, defined as
\begin{equation}
	T_{\rm  vir}\equiv\frac{G\mu m_{\rm H}M_{\rm  vir}}{3k_{\rm  B}r_{\rm  vir}},
\end{equation}
where $k_{\rm B}$ is the Boltzmann constant, $\mu$ is the mean molecular weight, and $m_{\rm H}$ is the mass of a hydrogen atom. 
The initial value for the temperature of gas, $T$, is assumed to be $T_{\rm vir}$.
A circular velocity, $v_{\rm c}$, is defined as
\begin{equation}
	v_{\rm  c}\equiv\left(\frac{GM_{\rm  vir}}{r_{\rm  vir}}\right)^\frac{1}{2},
\end{equation}
and we also define a rotation timescale, $t_{\rm cir}$, as
\begin{equation}
	t_{\rm  cir}\equiv\frac{2\pi r_{\rm disk}}{v_{\rm  c}}.
\end{equation}
Note that the rotation timescale of the gas disk, $t_{\rm cir}$, depend only on virialization redshift, $z_{\rm vir}$, as
\begin{equation}
	t_{\rm cir}=9.3\times10^{7}\ {\rm yr} \left(\frac{11}{1+z_{\rm vir}}\right)^{\frac{3}{2}}\left(\frac{18\pi^{2}}{\delta_{c}(z_{\rm vir})}\right)^{\frac{1}{2}}.
\end{equation}

We must estimate the number density of the hydrogen gas, $n_{\rm H}$, because it affects both the chemical reaction rate and the cooling rate. 
The cooling time of halo gas is much shorter than the Hubble timescale for the objects of interest in this paper ($T_{\rm vir}\gtrsim10000K$) \citep[e.g.][]{Hut02}.
It is widely understood that most $z\sim10$ galaxies were not clear disk galaxies; numerical simulations that proceed from cosmological initial conditions ($z\sim100-200$) clearly reveal that they possess highly irregular structures whose SF rates are not easily quantifiable, that filamentary accretion and frequent mergers are still churning the halo at this epoch, and that turbulent flows arise in the center of the halo that prevent coherent disks forming on the spatial scales of galaxies \citep{Joh08, Gre10, Wis10}. 
We assume that a significant fraction of baryons finally collapses to a disk in the dark matter halo potential for simplicity. 
%since  
In semi-analytic models, it is assumed that the cooled halo gas settles into the disk \citep[e.g.][]{Col00}. 
We make similar assumption in our model, but more detailed treatment for ${\rm H}_{2}$ formation, dust evolution and star formation in the gas disk. 
The radius of disk, $r_{\rm disk}$, is determined following \citet{Hir02}. 
We estimate the typical scale height, $H$, from hydrostatic equilibrium \citep{Sha88}
\begin{eqnarray}
	H &=& \sqrt{2}\frac{v_{\rm s}}{v_{\rm c}}r_{\rm disk} \nonumber\\
	&=& \left(\frac{2T}{3T_{\rm vir}}\right)^{\frac{1}{2}}r_{\rm disk}
	\label{eq:H}
\end{eqnarray}
for $H/r_{\rm disk}\le0.1$, otherwise, assume $H/r_{\rm disk}=0.1$, where $v_{s}=(k_{B}T/\mu m_{\rm H})^{\frac{1}{2}}$ is the isothermal sound velocity. 
Therefore, initial hydrogen density of disk, $n_{\rm H}$, is estimated as
\begin{equation}	
	n_{\rm H}=\frac{M_{\rm H}}{\pi r_{\rm  disk}^{2}2Hm_{\rm H}}.
\end{equation}
The initial mass of hydrogen in the galaxy, $M_{\rm H}$, is written as 
\begin{eqnarray}
	M_{\rm H}&=&M_{\rm gas}-M_{\rm He} \nonumber\\
	&=&M_{\rm gas}\frac{m_{\rm H}}{(m_{\rm  H}+m_{\rm He}y_{\rm He})} \nonumber\\
	&=&\frac{M_{\rm  vir}\Omega_{b}}{\Omega_{M}}\frac{m_{\rm H}}{(m_{\rm  H}+m_{\rm He}y_{\rm He})}
\end{eqnarray}
where $M_{\rm He}$ is the initial mass of helium in the galaxy, $M_{\rm gas}=M_{\rm H}+M_{\rm He}$, $m_{\rm He}$ is the mass of a helium atom and $y_{\rm He}$ is the helium abundance. 
We assume $y_{\rm He}=0.0972$ \citep{Omu00}. 
Note that since massive star ionizes surrounding gas and forms an expanding H II region \citep[e.g.][]{Wha04, Kit04}, we adopt gas density around SN progenitor, $n_{\rm SN}$, as being different from the hydrogen gas density in our one-zone galaxy model, $n_{\rm H}$. 
This is because the gas density around SN progenitor, $n_{\rm SN}$, is closely related with dust destruction process by reverse shocks driven by SNe, as shown in Section \ref{subsec:dustprod}. 

\subsection{Star formation law}
\label{subsec:sf} 
We expect that the SFR, $\Psi(t)$, is roughly proportional to $t_{\rm cir}^{-1}$, since a representative timescale of the dynamics of the gas disk is $t_{\rm cir}$. 
We assume that
\begin{equation}
	\Psi(t)=\frac{f_{{\rm  H}_{2}}(t)M_{\rm H}(t)}{t_{\rm  cir}(z_{\rm vir})},
	\label{SF}
\end{equation}
where $f_{\rm H_{2}}$ is the mass fraction of molecular hydrogen to the total gas. 
We should note that observationally \citet{Big08} find that ${\rm H}_{2}$ is converted into stars at a constant efficiency in nearby spirals and \citet{Gne09} show that the star formation recipe in galaxy formation simulation in which star formation occurs only in the molecular gas can reproduce the observational correlations between SFR and the total gas density. 

\subsection{Evolution of gas, stars, and metals}
\label{subsec:ism}
We calculate the time evolutions of the masses of hydrogen and helium gases, $M_{\rm gas}$, stars, $M_{\rm star}$, and metal of element $i$, $M_{{\rm m},i}$, in the galaxy by using the following equations
\begin{eqnarray}
	\frac{{\rm d}M_{\rm gas}(t)}{{\rm d}t}&=&-\Psi(t)\frac{M_{\rm gas}(t)}{M_{\rm ISM}(t)}+\overline{m_{\rm gas}}\gamma_{\rm SN}(t) \nonumber\\
	\frac{{\rm d}M_{\rm star}(t)}{{\rm d}t}&=&\Psi(t)-\overline{m_{\rm ejecta}}\gamma_{\rm SN}(t) \nonumber\\
	\frac{{\rm d}M_{{\rm m},i}}{{\rm d}t}&=&-\Psi(t)\frac{M_{{\rm m},i}(t)}{M_{\rm ISM}(t)}+\overline{m_{{\rm m},i}}\gamma_{\rm SN}(t) \nonumber\\
	&&
\end{eqnarray}
where $M_{\rm ISM}(t) = M_{\rm gas}(t)+\sum_{i}M_{{\rm m},i}(t)$, $\overline{m_{\rm ejecta}} = \overline{m_{\rm gas}}+\sum_{i}\overline{m_{{\rm m},i}}$ and $\overline{m_{\rm gas}}$ and $\overline{m_{{\rm m},i}}$ are the gas mass of hydrogen and helium and the metal mass of element $i$ in SN ejecta, respectively. 
The mass returning to the ISM, $m_{\rm ejecta}(m)$, $m_{\rm gas}(m)$ and $m_{{\rm m},i}(m)$ through a SN with progenitor mass, $m$, are taken from \citet{Ume02} in the case of $m=13$, $20$, $25$ and $30$ $M_{\odot}$. 
$\overline{m_{\rm gas}}$, $\overline{m_{\rm ejecta}}$ and $\overline{m_{m,i}}$ are IMF-averaged $m_{\rm gas}(m)$, $m_{\rm ejecta}(m)$ and $m_{\rm m,i}(m)$, respectively.
Note that metals consist of not only heavy elements in gas phase but also those in dust grains. 

\subsection{Chemistry and cooling}
\label{subsec:chemcool}
We follow the time evolution of molecular mass fraction, $f_{{\rm H}_{2}}$, ionization degree, $x$, and gas temperature, $T$. 
We define the molecular fraction of hydrogen as 
\begin{equation}
	f_{{\rm H}_{2}}\equiv\frac{2n_{{\rm H}_{2}}}{n_{\rm H}},
\end{equation}
where $n_{{\rm H}_{2}}$ and $n_{\rm H}$ are the number densities of molecular hydrogen and hydrogen nuclei, respectively.
The molecular fraction is very important in our models, because it determines the final cooling rate of low-metallicity gas. 
The metal-free gas evolution with chemical reactions and cooling is studied using the model by \citet{Teg97}, \citet{Hut02} and \citet{Hir02}. 
We summarize chemical reactions considered in this paper and their rate coefficients ($R_{n};n=1,\dots,11$) in Table \ref{tab:rea}. 
The equations are based on \citet{Hir02}, but we include the effect of the dust size distribution on ${\rm H}_{2}$ formation and the metal-line cooling process.  

The time evolution of the ionization degree is described as
\begin{equation}
	\frac{{\rm  d}x}{{\rm  d}t}=xf_{0}R_{1}n_{\rm  H}-x^{2}R_{2}n_{\rm  H}+\Gamma_{12}f_{0},
\end{equation}
where $f_{0}=1-x-f_{{\rm H}_{2}}$ is the neutral fraction of hydrogen.
The terms on the right-hand side are the rates of collisional ionization, recombination and photoionization. 
Next, the time evolution of the molecular fraction is written as
\begin{eqnarray}
	\frac{{\rm  d}f_{{\rm  H}_{2}}}{{\rm  d}t}&=&\left[\frac{{\rm  d}f_{{\rm  H}_{2}}}{{\rm  d}t}\right]_{\rm gas}+\left[\frac{{\rm  d}f_{{\rm  H}_{2}}}{{\rm  d}t}\right]_{\rm dust}+\left[\frac{{\rm  d}f_{{\rm  H}_{2}}}{{\rm  d}t}\right]_{\rm dest}  \nonumber\\
	&&+\left[\frac{{\rm  d}f_{{\rm  H}_{2}}}{{\rm  d}t}\right]_{\rm UV}+\left[\frac{{\rm  d}f_{{\rm  H}_{2}}}{{\rm  d}t}\right]_{\rm star}, 
\end{eqnarray}
where the terms on the right-hand side are the ${\rm H}_{2}$ formation rate in gas phase, the ${\rm H}_{2}$ formation rate on dust grains, the destruction rate in gas phase, and the destruction rate by UV photons, and the decreasing rate by star formation, respectively.
These terms are given by 
\begin{eqnarray}
	\left[\frac{{\rm  d}f_{{\rm  H}_{2}}}{{\rm  d}t}\right]_{\rm gas}&=&2f_{0}^{2}xn_{\rm  H}(R_{{\rm  eff},1}+R_{{\rm  eff},2}), \nonumber\\
	\left[\frac{{\rm  d}f_{{\rm  H}_{2}}}{{\rm  d}t}\right]_{\rm dust}&=&2R_{\rm  dust}{\cal D}n_{\rm  H}f_{0}, \nonumber\\
	\left[\frac{{\rm  d}f_{{\rm  H}_{2}}}{{\rm  d}t}\right]_{\rm dest}&=&-f_{{\rm  H}_{2}}n_{\rm  H}(x^{2}R_{{\rm  eff},3}+f_{0}R_{10}+xR_{11}), \nonumber\\
	\left[\frac{{\rm  d}f_{{\rm  H}_{2}}}{{\rm  d}t}\right]_{\rm UV}&=&-\Gamma_{13}f_{{\rm  H}_{2}}, \nonumber
\end{eqnarray}
and
\begin{eqnarray}
	\left[\frac{{\rm  d}f_{{\rm  H}_{2}}}{{\rm  d}t}\right]_{\rm star}&=&-(1-f_{{\rm H}_{2}})\Psi(t)\frac{1}{M_{\rm ISM}(t)}. \nonumber\\
	\label{fHtwoformation}
\end{eqnarray}
The effective formation rates of ${\rm H}_{2}$ including the effect of destruction rate of ${\rm H}^{-}$ and ${\rm H}_{2}^{+}$ are
\begin{equation}
	R_{{\rm eff},1}\equiv\frac{R_{3}R_{4}}{f_{0}R_{4}+xR_{5}+\Gamma_{14}/n_{\rm H}},
\end{equation}
and 
\begin{equation}
	R_{{\rm eff},2}\equiv\frac{R_{6}R_{7}}{f_{0}R_{7}+xR_{8}+\Gamma_{15}/n_{\rm H}},
\end{equation}
respectively, and the destruction of ${\rm H}_{2}^{+}$ due to ${\rm H}^{-}$ collision is
\begin{equation}
	R_{{\rm eff},3}\equiv\frac{R_{8}R_{9}}{f_{0}R_{7}+xR_{8}+\Gamma_{15}/n_{\rm H}}.
\end{equation}
We will give the dust-to-gas mass ratio, ${\cal D}$, and the production rate of molecular hydrogen via dust surface reaction, $R_{\rm dust}$, in Section \ref{subsec:formmol} and reaction rates of photo-process, $\Gamma_{n}(n=12,\dots,15)$, in Section \ref{subsec:rad}. 

At temperature $<10^{4}\ {\rm K}$, the main coolant is molecular hydrogen in low-metallicity gas. 
The cooling rate for molecular hydrogen, $\Lambda_{{\rm H}_{2}}$, over the range $10{\rm K}\le T\le10^{4}{\rm K}$ is given by \citep{Gal98}
\begin{eqnarray}
	&&\log_{10}\left(\frac{\Lambda_{{\rm  H}_{2}}(T)}{n_{\rm  H}n_{{\rm  H}_{2}}\ {\rm  erg}\ {\rm  cm}^{3}\ {\rm  s}^{-1}}\right)\nonumber \\
	&&\quad=-103.0+97.59T_{\rm  log}-48.05T_{\rm  log}^{2} \nonumber\\
	&&\quad+10.80T_{\rm  log}^{3}-0.9032T_{\rm  log}^{4}, 
\end{eqnarray}
where $T_{\rm log}\equiv\log_{10}(T/{\rm K})$. 
\citet{Glo08} have recently given the ${\rm H}_{2}$ cooling rates which include ${\rm H}-{\rm H}_{2}$ collision and ${\rm H}_{2}-{\rm H}_{2}$ collision pathways, while the \citet{Gal98} rates include only ${\rm H}-{\rm H}_{2}$ collisions. 
We assume that the lower-limit of gas temperature is the CMB temperature. 

At temperature $T\gtrsim10^{4}\ {\rm K}$, collisional excitation, $\Lambda_{\rm H,ce}$, and (less importantly) ionization of atomic hydrogen, $\Lambda_{\rm H,ci}$, are more dominant cooling process than molecular hydrogen cooling and are given by \citep{Hai96} 
\begin{equation}
	\frac{\Lambda_{\rm  H,ce}(T)}{n_{{\rm  e}^{-}}n_{\rm  H}\ {\rm  erg}\ {\rm  cm}^{3}\ {\rm  s}^{-1}}=7.50\times10^{-19}\frac{1}{1+T_{5}^{\frac{1}{2}}}\ \exp^{-\frac{1.183}{T_{5}}}
\end{equation}
and
\begin{equation}
	\frac{\Lambda_{\rm  H,ci}(T)}{n_{{\rm  e}^{-}}n_{\rm H}\ {\rm  erg}\ {\rm  cm}^{3}\ {\rm  s}^{-1}}=4.02\times10^{-19}\frac{T_{5}^{\frac{1}{2}}}{1+T_{5}^{\frac{1}{2}}}\ \exp^{-\frac{1.578}{T_{5}}}
\end{equation}
respectively, where $T_{5}$ is gas temperature in units of $10^{5}\ {\rm K}$. 

We consider fine-structure cooling by ${\rm C}_{\rm I}$, ${\rm C}_{\rm II}$ and ${\rm O}_{\rm I}$, which dominates the thermal evolution for number density of gas of interest in this paper \citep{Omu05}. 
The related parameters of transitions are given in \citet{Hol89}.
\begin{center}
\begin{table*}
\caption{Reaction rates needed to calculate the abundance of ${\rm H}_{2}$. 
The unit of the gas temperature $T$ is ${\rm K}$ unless otherwise stated. 
Reference: 1, \citet{Omu00}; 2, \citet{Gal98}}
\begingroup
\renewcommand{\arraystretch}{0.7}
\begin{tabular*}{17.5cm}{@{\extracolsep{\fill}}cllc}\hline
No. & Reaction & Rate [cm$^3$ s$^{-1}$] & Ref. \\ \hline
1 & ${\rm H+e^{-}\longrightarrow H^++2e^-}$ & $\exp [-32.71+13.54\ln (T({\rm eV}))$ & 1\\
  & & $~-5.739(\ln (T({\rm eV})))^2+1.563(\ln (T({\rm eV})))^3$ & \\
  & & $~-0.2877(\ln (T({\rm eV})))^4+3.483\times10^{-2}(\ln (T({\rm eV})))^5$ & \\
  & & $~-2.632\times 10^{-3}(\ln (T({\rm eV})))^6$ & \\
  & & $~+1.120\times 10^{-4}(\ln (T({\rm eV})))^7$ & \\
  & & $~-2.039\times 10^{-6}(\ln (T({\rm eV})))^8]$ &  \\
2 & ${\rm H^++e^-\longrightarrow H+\gamma}$ & $\exp [-28.61-0.7241(\ln (T({\rm eV})))$ & 1 \\
  & & $~-2.026\times 10^{-2}(\ln(T({\rm eV})))^2$ & \\
  & & $~-2.381\times 10^{-3}(\ln (T({\rm eV})))^3$ & \\
  & & $~-3.213\times 10^{-4}(\ln (T({\rm eV})))^4$ & \\
  & & $~-1.422\times 10^{-5}(\ln (T({\rm eV})))^5$ & \\
  & & $~+4.989\times 10^{-6}(\ln (T({\rm eV})))^6$ & \\
  & & $~+5.756\times 10^{-7}(\ln (T({\rm eV})))^7$ & \\
  & & $~-1.857\times 10^{-8}(\ln (T({\rm eV})))^8$ & \\
  & & $~-3.071\times 10^{-9}(\ln (T({\rm eV})))^9]$ & \\
3 & ${\rm H+e^{-}\longrightarrow H^-+\gamma}$ & $1.4\times 10^{-18}T^{0.928}\exp (-T/1.62\times 10^4)$ & 1 \\ 
4 & ${\rm H^-+H\longrightarrow H_2+e^-}$ & $4.0\times 10^{-9}T^{-0.17}~(T>300)$;  & 1 \\
  & & $1.5\times 10^{-9}~(T<300)$ & \\
5 & ${\rm H^-+H^+\longrightarrow 2H}$ & $5.7\times 10^{-6}T^{-1/2}+6.3\times 10^{-8}$ & 1 \\
  & & $~-9.2\times 10^{-11}T^{1/2}+4.4\times 10^{-13}T$ & \\
6 & ${\rm H+H^+\longrightarrow H_2^++\gamma}$ & ${\rm dex}[-19.38-1.523\log_{10}T$ & 1 \\
  & &$~+1.118(\log_{10}T)^2-0.1269(\log_{10}T)^3]$ & \\
7 & ${\rm H_2^++H\longrightarrow H_2+H^+}$ & $6.4\times 10^{-10}$ & 1 \\
8 & ${\rm H_2^++e^-\longrightarrow 2H}$ & $2.0\times 10^{-7}T^{-1/2}$ & 1 \\
9 & ${\rm H_2+H^+\longrightarrow H_2^++H}$ & $3.0\times 10^{-10}\exp (-21050/T)$ $(T<10^4)$ & 2 \\
  & & $1.5\times 10^{-10}\exp (-14000/T)$ $(T>10^4)$ & \\
10 & ${\rm H_2+H\longrightarrow 3H}$ & $k_{\rm H}^{1-a}k_{\rm L}^a$ & 1 \\
  & & $~k_{\rm L}=1.12\times 10^{-10}\exp (-7.035\times 10^4/T)$ & \\
  & & $~k_{\rm H}=6.5\times 10^{-7}T^{-1/2}$ & \\
  & & $~~\times\exp (-5.2\times 10^4/T)[1-\exp (-6000/T)]$ & \\
  & & $~a=4.0-0.416\log_{10}(T/10^4)-0.327(\log_{10}(T/10^4))^2$ & \\
11 & ${\rm H_2+e^-\longrightarrow 2H+e^-}$ & $4.4\times 10^{-10}T^{0.35}\exp (-1.02\times 10^5/T)$ & 1 \\
 & & & \\
dust & ${\rm H+H+\mbox{grain}\longrightarrow H_2+\mbox{grain}}$ & see Section \ref{subsec:formmol} & \\
& & & \\
\hline
\end{tabular*}
\endgroup
\label{tab:rea}
\end{table*}
\end{center}

\subsection{Formation of molecular hydrogen on dust grains}
\label{subsec:formmol}
The increasing rate of molecular fraction via dust surface reaction is estimated as
\begin{eqnarray}
	\left[\frac{{\rm d}f_{{\rm H}_{2}}}{{\rm d}t}\right]_{\rm dust}&=&2R_{\rm dust}{\cal D}n_{\rm H}f_{0} \nonumber\\
	&=&\sum_{j}\int_{0}^{\infty} f_{0}f_{j}(a)\pi a^{2}\bar{v}S {\rm d}a \nonumber\\
	&&
\end{eqnarray}
where $\bar{v}$ is the mean thermal speed of hydrogen and $S$ is the sticking efficiency of hydrogen atoms. 
We assume that the gas follows a Maxwellian distribution so that thermal speed is given by \citep{Kru08} 
\begin{equation}
	\bar{v}=\left(\frac{8}{\pi}\frac{k_{\rm B}T}{m_{\rm H}}\right)^{\frac{1}{2}}.
\end{equation}
Here, we define the dust-to-gas mass ratio, ${\cal D}$, as
\begin{equation}
	{\cal D}\equiv\sum_{j}\int_{0}^{\infty}\frac{4\pi a^{3}\rho_{j}f_{j}(a)}{3n_{\rm H}m_{\rm H}}{\rm d}a
\end{equation}
The reaction rate of the ${\rm H}_{2}$ formation on grains, $R_{\rm dust}$, can be estimated as
\begin{equation}
	R_{\rm dust}(a){\cal D}=\sum_{j}\int_{0}^{\infty}\left(\frac{3m_{\rm H}\bar{v}S}{8a\rho_{j}}\right)\left(\frac{4\pi a^{3}\rho_{j}f_{j}(a)}{3n_{\rm H}m_{\rm H}}\right){\rm d}a.
\label{eq:dustreaction}
\end{equation}
We adopt $S=0.2$ for $T<300\ {\rm K}$ and $S=0$ for $T>300\ {\rm K}$ \citep{Hir02}. 
\subsection{Radiative properties}
\label{subsec:rad}
We follow photo-processes in chemical reaction and heating processes in thermal evolution by using the models developed by \citet{Kit00} and improved in \citet{Hir02}. 
The intrinsic luminosity is assumed to be equal to the total luminosity of OB stars whose mass is larger than $3\ M_{\odot}$ \citep{Cox00}
\begin{equation}
	L_{\rm UV,0}(t)=\int_{3\ M_{\odot}}^{\infty}{\rm d}m\int_{0}^{\tau_{m}}{\rm d}t^{\prime}\ L(m)\phi(m)\Psi(t-t^{\prime}),
\end{equation}
where $L(m)$ is the stellar luminosity as a function of stellar mass $m$. 
For $L(m)$, we adopt the model of zero-metallicity stars without mass loss in \citet{Sch02}. 
We assume the spectrum of the incident UV radiation from stars is a power law with an index $\alpha$ :
\begin{equation}
I_{\rm UV}(\nu)=I_{0}(\nu_{\rm HI})\left(\frac{\nu}{\nu_{\rm HI}}\right)^{-\alpha}
\end{equation}
where $\nu$ is the frequency of photons and $I_{0}(\nu_{\rm HI})$ is the intensity at the ionization threshold frequency of neutral hydrogen $\nu_{\rm HI}=3.3\times10^{15}\ {\rm Hz}$. 
In this paper, we simply set $\alpha=5$ according to \citet{Hir02}. 
The normalization of the intensity is determined by 
\begin{equation}
\frac{L_{\rm UV,0}\exp(-\tau_{\rm disk})}{4\pi r_{\rm disk}^{2}}=\int_{\nu_{\rm min}}^{\infty}I_{\rm UV}(\nu){\rm d}\nu,
\end{equation}
where $\nu_{\rm min}$ is the minimum frequency where OB stars dominate the radiative energy of star-forming galaxies, and $\tau_{\rm disk}$ is the typical dust optical depth in the disk. 
We assume that $\nu_{\rm min}=10^{15}\ {\rm Hz}$. 
This typical optical depth can be simply estimated as 
\begin{equation}
	\tau_{\rm disk}=r_{\rm disk}\sum_{j}\int_{0}^{\infty}\pi a^{2}f_{j}(a){\rm d}a
\end{equation}
by assuming that extinction efficiency of dust is unity in UV. 
We calculate $\Gamma_{12}$, $\Gamma_{14}$ and $\Gamma_{15}$ from Equation (A20) of \citet{Kit00}  and heating rate from Equation (A21) of \citet{Kit00}. 
We summarize the cross section for the photo-process in Table \ref{tab:pho}. 
The H$_{2}$ photodissociation cross-section is given by \citet{Abe97}.
However, if the ${\rm H}_{2}$ column density becomes larger than $10^{14}\ {\rm cm}^{-2}$, self-shielding effects become important \citep{Dra96}. 
Therefore, ${\rm H}_{2}$ dissociation rate, $\Gamma_{13}$, is given by \citep{Hir02}
\begin{eqnarray}
	\Gamma_{13}&=&(4\pi)1.1\times10^{8}I_{\rm UV}(3.1\times10^{15}\ {\rm Hz}) \nonumber\\
	&&\times\left(\frac{n_{\rm H}f_{{\rm H}_{2}}r_{\rm disk}}{10^{14}\ {\rm cm}^{-2}}\right)^{-0.75}\ {\rm s}^{-1}, 
\label{eq:gamma13}
\end{eqnarray}
where $I_{\rm UV}(3.1\times10^{15}\ {\rm Hz})$ is in the Lyman-Werner band. 
We should note that adoption of $r_{\rm disk}$ in Equation (\ref{eq:gamma13}) gives an extreme upper bound to the self-shielding, so we may overestimate self-shielding to internal Lyman-Werner photons by ${\rm H}_{2}$. 
In this paper, we focus on the effects of dust on the protogalaxy, so for simplicity, we set an extreme upper bound to the self-shielding. 

We do not consider the Lyman-Werner background, since we concentrate in evolution of atomic line cooling halos ($M_{\rm vir} > 10^{8}\ M_{\odot}$ in $z < 10$) in which destruction of molecular hydrogen by the Lyman-Werner background is less efficient \citep{O'S08, Sus08, Wis08, Wis09}. 
However, $10^{8}-10^{9}\ {M_{\odot}}$ halos are not immune to the Lyman-Werner background, just self-shielded at their very centers. 
Not all the baryons will be protected from external photodissociating flux and this will affect ${\rm H}_{2}$ production on dust outside the center of halo. 
In the lower mass halos, the Lyman-Werner background may be effective to dissociate ${\rm H}_{2}$ molecule \citep{Mac01, Mac03, Yos03, Sus07, Wis07, O'S08}. 
\begin{center}
\begin{table*}
\caption{Cross-sections for photoionization and photodissociation process, where the frequency, $\nu$, is in units of ${\rm Hz}$. Reference: 1, \citet{Kit00}; 2, \citet{Abe97}; 3, \citet{Teg97} }
\begingroup
\renewcommand{\arraystretch}{0.7}
\begin{tabular*}{17.5cm}{@{\extracolsep{\fill}}clllc}\hline
No. & Reaction & cross section & $\nu$ range & Ref. \\
& & (cm$^2$) & (Hz) & \\ \hline
12 & ${\rm H+\gamma\longrightarrow H^++e^-}$ & $6.30\times 10^{-18}(\nu /3.3\times 10^{15})^{-3.0}$ & $\nu >3.3\times 10^{15}$ & 1 \\
13 & ${\rm H_2+\gamma\longrightarrow H_2^*\longrightarrow 2H}$ & see equation (\ref{eq:gamma13}) & & 2 \\
14 & ${\rm H^-+\gamma\longrightarrow H+e^-}$ & $3.486\times 10^{-16}(x-1)^{3/2}/x^{3.11}$ &$\nu >1.8\times 10^{14}$ & 3 \\
  & & ~($x\equiv \nu /1.8\times 10^{14}$) &  &  \\
15 & ${\rm H_2^++\gamma\longrightarrow H+H^+}$ & $7.401\times 10^{-18}$ & $~\nu >6.4\times 10^{14}$ & 3 \\
  & & $~{\rm dex}(-x^2-0.0302x^3-0.0158x^4)$ & & \\
  & & ($x\equiv 2.762\ln (\nu /2.7\times 10^{15}$) & & \\
\hline
\end{tabular*}
\endgroup
\label{tab:pho}
\end{table*}
\end{center}

\section{Results}
\label{sec:resultevo}
Our fiducial model assumes $M_{\rm vir}=10^{9}\ M_{\odot}$ and $z_{\rm vir}=10$, and includes the dust destruction model by both reverse shocks and forward shocks with the ISM density around the SN progenitor, $n_{\rm SN}=1\ {\rm cm}^{-3}$ (see Table \ref{tab:model} for a summary of our models). 
This dark matter halo forms from a $2.5\ \sigma$ density fluctuation. 
We stop the calculation at $z=5$. 
It corresponds to the galaxy age of $\sim 0.8\ {\rm Gyr}$ that is before SN II to be dominant source of dust grains. 
The initial mass of gas is $M_{\rm gas} = 1.7\times10^{8}\ M_{\odot}$ and the dynamical timescale of circular motion of the gas disk is $t_{\rm cir}(z_{\rm vir}=10)=9.5\times10^7\ {\rm yr}$. 
Note that galaxies with $M_{\rm vir}\sim10^{9}\ M_{\odot}$ play a critical role in the cosmic reionization, since in the relevant redshift range for cosmological reionization, $z=6-15$, most of reionization radiation is expected to come from galaxies with masses less than $\sim10^{9.5} M_{\odot}$ \citep{Wis09}. 
\begin{table*}
\caption{the dust destruction model and the main parameters.}
\begingroup
\renewcommand{\arraystretch}{1.0}
\begin{tabular*}{14cm}{@{\extracolsep{\fill}}clcc}\hline
model & dust destruction & $n_{\rm SN} ({\rm cm}^{-3})$ & $M_{\rm vir}$ ($M_{\odot}$) \\ \hline
C1m9 & no destruction & -- & $10^{9}$ \\
B1m9 & forward shocks & 1 & $10^{9}$ \\
A1m9  (fiducial) & forward and reverse shocks & $1$ & $10^{9}$ \\
A0.03m9 & forward and reverse shocks & 0.03 & $10^{9}$ \\
A0.1m9 & forward and reverse shocks & 0.1 & $10^{9}$ \\
A0.3m9 & forward and reverse shocks & 0.3 & $10^{9}$ \\
A3m9 & forward and reverse shocks & 3 & $10^{9}$ \\
A10m9 & forward and reverse shocks & 10 & $10^{9}$ \\
A0.1m8 & forward and reverse shocks & 0.1 & $10^{8}$ \\
A0.1m10 & forward and reverse shocks & 0.1 & $10^{10}$ \\
A0.1m11 & forward and reverse shocks & 0.1 & $10^{11}$ \\
A1m8 & forward and reverse shocks & 1 & $10^{8}$ \\
A1m10 & forward and reverse shocks & 1 & $10^{10}$ \\
A1m11 & forward and reverse shocks & 1 & $10^{11}$ \\
A10m8 & forward and reverse shocks & 10 & $10^{8}$ \\
A10m10 & forward and reverse shocks & 10 & $10^{10}$ \\
A10m11 & forward and reverse shocks & 10 & $10^{11}$ \\
\hline
\end{tabular*}
\endgroup
\label{tab:model}
\end{table*}

\subsection{The dust destruction}
\label{res:dustdest}
We first show the evolution of a galaxy with $M_{\rm vir}=10^{9}\ M_{\odot}$ and $z_{\rm vir}=10$ for various dust destruction models. 
To clarify the dust destruction effects on galaxy evolution, we first show the result of the model without reverse shocks and forward shocks (model C1m9), then compare the results of the models with only forward shocks (model B1m9) and with both forward and reverse shocks (model A1m9) to the model without both shocks (model C1m9). 
\begin{figure}[!h]
\epsscale{1.00}
\plotone{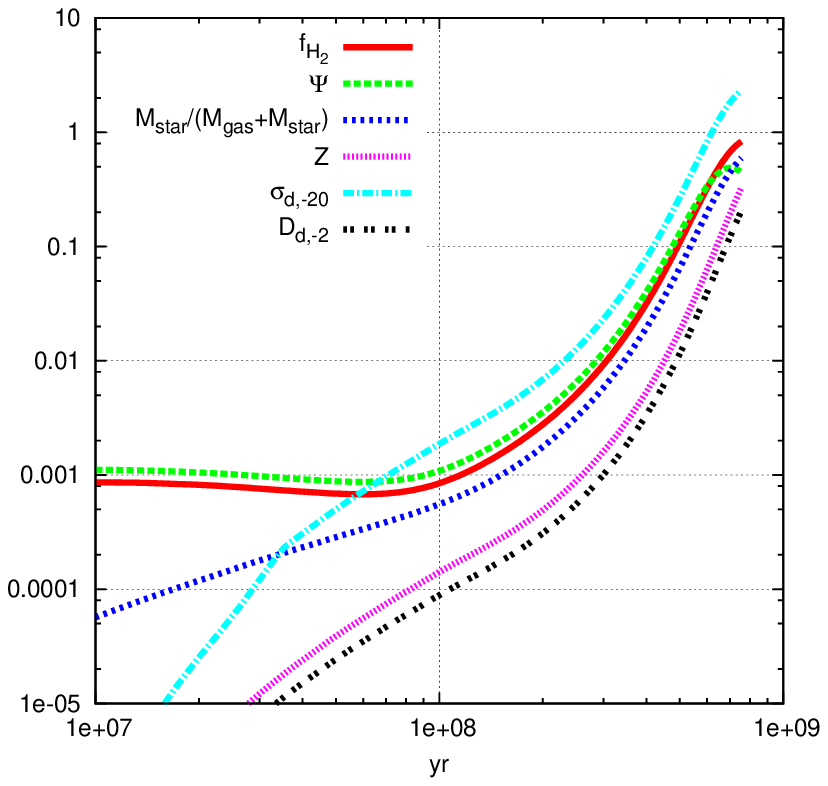}
\caption{Time evolution for model C1m9 in which dust destruction by the reverse shocks and the forward shocks are not considered. We show molecular fraction, $f_{{\rm H}_{2}}$, SFR in units of $M_{\odot}\ {\rm yr}^{-1}$, $\Psi$, star-to-gas mass ratio, $M_{\rm star}/(M_{\rm gas}+M_{\rm star})$, metallicity, $Z$, in units of $Z_{\odot}$, total dust cross-section per unit volume, $\sigma_{{\rm d},-20}$, in unit of $10^{-20}\ {\rm cm}^{-1}$, and dust-to-gas mass ratio, $D_{{\rm d},-2}$, in units of $10^{-2}$. 
The horizontal axis shows the age of galaxy in unit of ${\rm yr}$ from the formation redshift, $z_{\rm vir}=10$.}
\label{0_6_2}
\end{figure}

In Figure \ref{0_6_2}, we show the evolution of various quantities without dust destruction (model C1m9). 
The figure shows the time evolution of the molecular fraction, $f_{{\rm H}_{2}}$, the SFR in units of $M_{\odot}\ {\rm yr}^{-1}$, $\Psi$, the stellar mass fraction, $M_{\rm star}/(M_{\rm gas}+M_{\rm star})$, the metallicity, $Z$, in units of $Z_{\odot}$, total dust cross-section per unit volume, $\sigma_{{\rm d},-20}$, in unit of $10^{-20}\ {\rm cm}^{-1}$, and the dust-to-gas mass ratio, $D_{{\rm d},-2}$, in units of $10^{-2}$. 
The definition of $D_{{\rm d},-2}$ is convenient for comparison with the MW value of dust-to-gas mass ratio. 
In the MW the dust-to-gas mass ratio is $0.5\times10^{-2}$ in the diffuse ISM \citep{Dra09} and $0.9\times10^{-2}$ in molecular clouds \citep{Pol94}. 

The molecular fraction, $f_{{\rm H}_2}$, is very important, since it determines SFR and controls galaxy evolution. 
The molecular fraction reaches $f_{{\rm H}_{2}}\sim1\times10^{-3}$ around $t\sim10^{7}\ {\rm yr}$. 
This results is robust for all models, since in this stage ${\rm H}_{2}$ formation in the gas phase is dominant over that on dust grains \citep{Teg97, Hir02}. 
The gas temperature rapidly drops below $200\ {\rm K}$ before $10^{7}\ {\rm yr}$. 
Then, the molecular fraction rapidly increases from $t\sim10^{8}\ {\rm yr}$ and reaches $\sim0.83$ at the galaxy age of $\sim0.8\ {\rm Gyr}$ ($z=5$). 
This is due to the enhancement of ${\rm H}_{2}$ formation on dust grains by increase of $\sigma_{{\rm d},-20}$. 
For $t\gtrsim\ 3\times10^{7}\ {\rm yr}$, $\sigma_{{\rm d},-20}\gtrsim1.1\times10^{-4}$ and $\left[{\rm d}f_{{\rm H}_{2}}/{\rm d}t\right]_{\rm dust}$ exceeds $\left[{\rm d}f_{{\rm H}_{2}}/{\rm d}t\right]_{\rm star}$. 
For $\sigma_{{\rm d},-20}\gtrsim0.001$, the increase of molecular fraction enhances the star formation. 
The cycle of the ${\rm H}_{2}$ formation on dust, the star formation and the dust formation by SNe, significantly accelerates galaxy evolution, such as rapid increase of the stellar mass fraction, $M_{\rm star}/(M_{\rm gas}+M_{\rm star})$. 
At the galaxy age $\sim0.8\times10^{9}\ {\rm yr}$, the stellar mass fraction goes up to $M_{\rm star}/(M_{\rm gas}+M_{\rm star})\sim0.60$. 
The SFR, $\Psi(t)$, decreases from the time when $M_{\rm star}/(M_{\rm gas}+M_{\rm star})\sim0.45$, since gas mass decreases significantly. 
The active star formation causes the formation of dust grains and metals. 
At $t\sim0.8\ {\rm Gyr}$, the total dust cross-section, $\sigma_{{\rm d},-20}$, goes up to $2.3$ and the metallicity, $Z$, goes up to $3.3\times10^{-1}\ Z_{\odot}$. 
\begin{figure}[h]
\epsscale{1.00}
\plotone{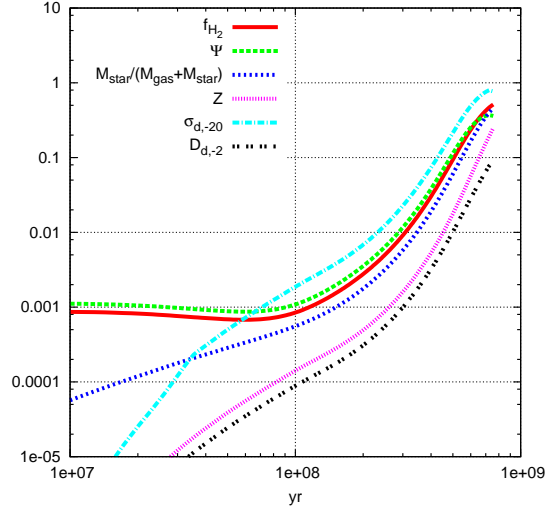}
\caption{Same as in Figure \ref{0_6_2} but for model B1m9 in which the dust destruction only by the forward shocks is considered. }
\label{0_6_1}
\end{figure}

Figure \ref{0_6_1} shows the results of the galaxy model with the dust destruction by only the forward shocks (model B1m9) to illustrate the effects of dust destruction by forward shocks on the galaxy evolution. 
In this model $n_{\rm SN}=1.0\ {\rm cm^{-3}}$. 
The dust destruction by forward shocks slightly affects the dust-to-gas mass ratio, $D_{{\rm d},-2}$, after the galaxy age of $\sim5\times10^{8}\ {\rm yr}$. 
This is because the destruction by forward shocks is roughly proportional to the dust-to-gas mass ratio (see Equation (\ref{dustformulation})), and dust grains are destroyed significantly for $D_{{\rm d},-2}\gtrsim0.1$ in this case. 
The SFR decreases from the time when $M_{\rm star}/(M_{\rm gas}+M_{\rm star})\sim0.4$. 
At $\sim0.8\ {\rm Gyr}$, the molecular fraction reaches $f_{{\rm H}_{2}}\sim0.51$ and the stellar mass fraction reaches $M_{\rm star}/(M_{\rm gas}+M_{\rm star})\sim0.47$. 
\begin{figure}[!h]
\epsscale{1.00}
\plotone{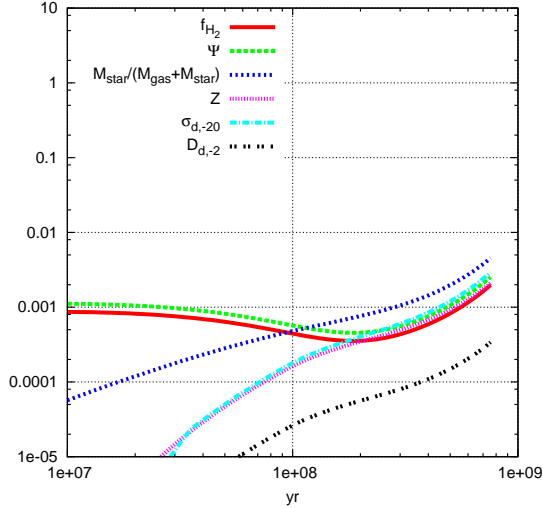}
\caption{Same as in Figure \ref{0_6_2} but for model A1m9, in which the destruction by reverse and forward shocks in the case of $n_{\rm SN}=1\ {\rm cm}^{-3}$} is considered. 
\label{0_5_1}
\end{figure}

In Figure \ref{0_5_1}, we show the results of our fiducial model A1m9 in which we include the dust destruction by both reverse and forward shocks in the case of $n_{\rm SN}=1\ {\rm cm^{-3}}$. 
The molecular fraction reaches to $f_{{\rm H}_{2}}\sim1\times10^{-3}$ around $t\sim10^{7}\ {\rm yr}$. 
This is similar to C1m9 and B1m9. 
After $t\gtrsim10^{7}\ {\rm yr}$, the molecular fraction evolution is quite different from models C1m9 and B1m9. 
The molecular fraction declines slowly until $t\sim2\times10^{8}\ {\rm yr}$. 
After $\sim2\times10^{8}\ {\rm yr}$, the molecular fraction increases slowly with increase of the dust mass. 
This is due to the ${\rm H}_{2}$ formation on the dust grains. 
The molecular fraction reaches only $\sim2.0\times10^{-3}$ at $t\sim0.8\ {\rm Gyr}$. 
This is because dust destruction by reverse shocks is very effective and hence results in suppression of ${\rm H}_{2}$ formation on dust grains. 
On the other hand, forward shocks hardly affect the evolution of dust size and dust mass, since the destruction of forward shocks can change dust mass only for large dust-to-gas mass ratio, $D_{{\rm d},-2}\gtrsim10^{-1}$. 
At $t\sim0.8\ {\rm Gyr}$, the stellar mass fraction reaches only $M_{\rm star}/(M_{\rm gas}+M_{\rm star})\sim4.5\times10^{-3}$, which is much less than the model without reverse shocks shown in Figure \ref{0_6_2} (model C1m9) and Figure \ref{0_6_1} (model B1m9). 

We illustrate the difference in the ${\rm H}_{2}$ formation rate among models C1m9, B1m9, and A1m9 as follows. 
The ${\rm H}_{2}$ formation rate depends not only on the total dust mass but also on the dust size distribution. 
In models C1m9 and B1m9, the dust mass produced by a SN II without reverse shock is $\sum_{j}\int\overline{{\cal M}^{0}_{\rm d,j}}(a){\rm d}a=0.48\ M_{\odot}$ and in the A1m9, the dust mass injection into ISM through a reverse shock with $n_{\rm SN}=1.0\ {\rm cm}^{-3}$ is $\sum_{j}\int\overline{{\cal M}^{1.0}_{\rm d,j}}(a){\rm d}a=0.15\ M_{\odot}$. 
The ratio of the mean dust area to the mean dust volume is $\langle a^{2}\rangle/\langle a^{3}\rangle=1.5\times10^{5}\ {\rm cm}^{-1}$ before the reverse shock destruction. 
After the reverse shock destruction, $\langle a^{2}\rangle/\langle a^{3}\rangle=4.2\times10^{4}\ {\rm cm}^{-1}$. 
This is a measure of dust area per the dust volume and is also a measure of ${\rm H}_{2}$ formation rate of the dust surface. 
Small $\langle a^{2}\rangle/\langle a^{3}\rangle$ leads to a low ${\rm H}_{2}$ formation rate. 
This is the reason why model A1m9 shows smaller ${\rm H}_{2}$ fraction than B1m9 and C1m9. 
The dust destruction by reverse shocks changes not only the dust mass but also the grain size distribution, and as a result drastically suppresses star formation in the galaxy. 
\begin{figure}[h]
\epsscale{1.00}
\plotone{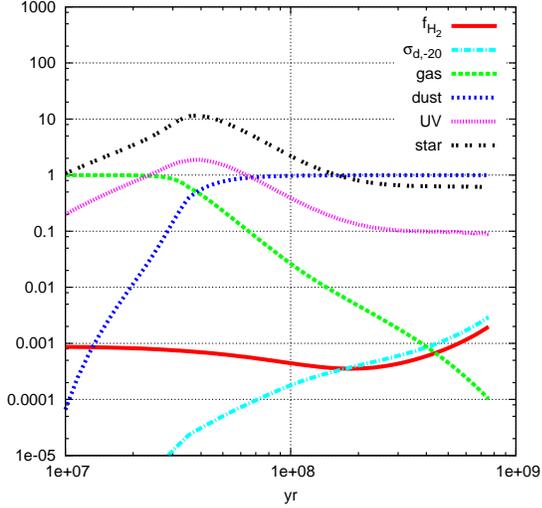}
\caption{Formation and destruction rates of ${\rm H}_{2}$ in model A1m9. 
The evolution of molecular formation rate in gas phase (gas), the molecular formation rate on dust grain (dust), the molecular destruction rate by UV photons (UV), and the molecular decreasing rate by the star formation (star), normalized to the total formation rate, are shown. 
The molecular formation on dust grains becomes dominant, compared with the other channels after $t\sim1.6\times10^8\ {\rm yr}$. 
The molecular fraction, $f_{{\rm H}_{2}}$, and total dust cross-section per unit volume in unit of $10^{-20}\ {\rm cm}^{-1}$, $\sigma_{{\rm d},-20}$, are the same as in Figure \ref{0_5_1}. }
\label{fHtwoform_5_1}
\end{figure}

In Figure \ref{fHtwoform_5_1}, we show the evolution of ${\rm H}_{2}$ formation rate in gas phase, $[{\rm d}f_{{\rm H}_2}/{\rm d}t]_{\rm gas}$, the ${\rm H}_{2}$ formation rate on dust grain, $[{\rm d}f_{{\rm H}_2}/{\rm d}t]_{\rm dust}$, the ${\rm H}_{2}$ destruction rate by UV photons, $[{\rm d}f_{{\rm H}_2}/{\rm d}t]_{\rm UV}$, and the ${\rm H}_{2}$ decreasing rate by the star formation, $[{\rm d}f_{{\rm H}_2}/{\rm d}t]_{\rm star}$, normalized to the total formation rate, $[{\rm d}f_{{\rm H}_2}/{\rm d}t]_{\rm gas}+[{\rm d}f_{{\rm H}_2}/{\rm d}t]_{\rm dust}$, in the model A1m9. 
At $t\sim4\times10^{7}\ {\rm yr}$, the ${\rm H}_{2}$ formation rate on dust grains exceeds the rate in gas phase. 
However, the ${\rm H}_{2}$ formation rate on dust grain is less than the ${\rm H}_{2}$ decreasing rate by star formation at this epoch. 
At $t\sim1.6\times10^{8}\ {\rm yr}$, the ${\rm H}_{2}$ formation rate on dust grains exceeds the ${\rm H}_{2}$ decreasing rate by star formation. 
The formation on dust grain becomes the dominant process among all of ${\rm H}_{2}$ formation and destruction processes. 
From this time when $\sigma_{{\rm d},-20}\gtrsim5.2\times10^{-5}$, molecular fraction, $f_{{\rm H}_{2}}$, starts to increase. 
The molecular destruction rate by UV photons does not exceed the molecular decreasing rate by the star formation after $t\sim4\times10^{7}\ {\rm yr}$. 
After $t\sim4\times10^{7}\ {\rm yr}$, $I_{\rm UV}(3.1\times10^{15}\ {\rm Hz})\sim2.0-4.8\times10^{-20}\ {\rm erg}\ {\rm s}^{-1}\ {\rm cm}^{-2}\ {\rm Hz}^{-1}\ {\rm ster}^{-1}$, which corresponds to $J_{21}=20-40$ where $J_{\rm 21}$ is in units of $I_{\rm UV}(3.1\times10^{15}\ {\rm Hz})=J_{21}\times10^{-21} \ {\rm erg}\ {\rm s}^{-1}\ {\rm cm}^{-2}\ {\rm Hz}^{-1}\ {\rm ster}^{-1}$. 
Note that $J_{\rm 21}$ in this model is higher than the values, $J_{\rm 21}\sim1$, for the Lyman-Werner background in the redshift of $5<z<10$ suggested in recent papers \citep[e.g.][]{Gre06}. 
The destruction process in gas phase does not affect the evolution of molecular fraction significantly after $t\gtrsim1\times10^{7}\ {\rm yr}$. 

\subsection{The ISM density around SN}
\begin{figure}[!h]
\epsscale{1.0}
\plotone{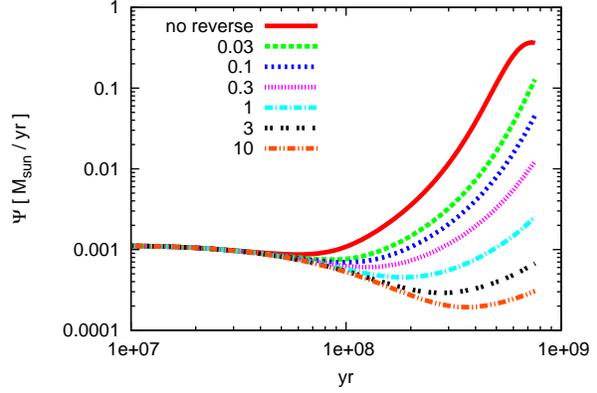}
\caption{Time evolution of SFR in unit of $M_{\odot}\ {\rm yr}^{-1}$ in the model with dust destruction for various $n_{\rm SN}=0.03$, $0.1$, $0.3$, $1.0$, $3.0$ and $10$ ${\rm cm}^{-3}$.
The values of $n_{\rm SN}$ are used in the panel as in Figure \ref{fig:prod}. 
The model without dust destruction by the reverse shocks is also shown. 
The horizontal axis shows the age of galaxy in unit of ${\rm yr}$ from the formation redshift, $z_{\rm vir}=10$. }
\label{SFevo2}
\end{figure}
\begin{figure}[!h]
\epsscale{1.0}
\plotone{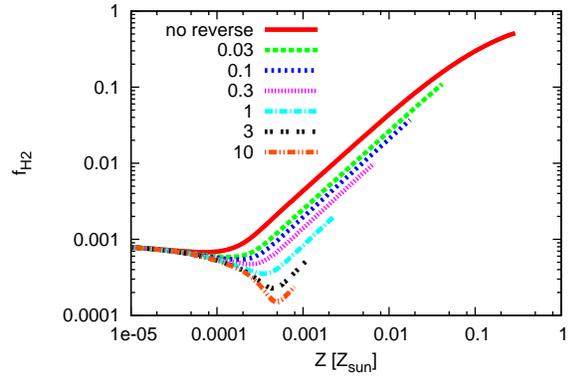}
\caption{Molecular faction evolution, $f_{{\rm H}_{2}}$, for the model of different densities around a SN II, $n_{\rm ISM, SN}=0.03$, $0.1$, $0.3$, $1.0$, $3.0$ and $10\ {\rm cm}^{-3}$, which are used in the panel as in Figure \ref{fig:prod}, and the model without reverse shocks. 
The horizontal axis shows the metallicity, $Z$, in unit of $Z_{\odot}$. }
\label{fHtwoevo}
\end{figure}
\begin{figure}[!h]
\epsscale{1.0}
\plotone{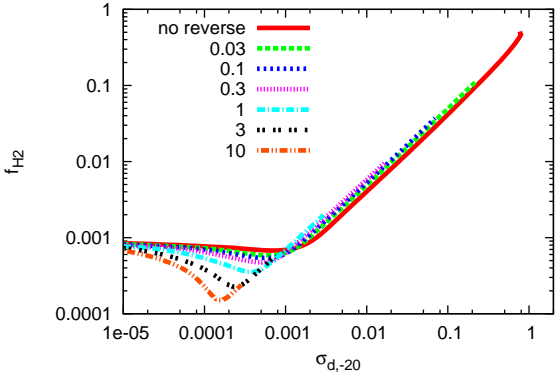}
\caption{Molecular faction evolution, $f_{{\rm H}_{2}}$, for the model of different densities around a SN II, $n_{\rm ISM, SN}=0.03$, $0.1$, $0.3$, $1.0$, $3.0$ and $10\ {\rm cm}^{-3}$, which are used in the panel as in Figure \ref{fig:prod}, and the model without reverse shocks. 
The horizontal axis shows the total dust cross-section per unit volume, $\sigma_{{\rm d},-20}$, in unit of $10^{-20}\ {\rm cm}^{-1}$. }
\label{fHtwoevo2}
\end{figure}

The dependence of time evolution of SFR on the ISM density around SN is presented in Figure \ref{SFevo2}. 
We can see that after $t\sim5\times10^{7}\ {\rm yr}$ higher density around SNe progenitors results in lower molecular fraction, and hence lower star formation efficiency. 
The SFR, $\Psi(t)$, is independent of $n_{\rm SN}$ before $t\sim5\times10^{7}\ {\rm yr}$, because ${\rm H}_{2}$ forms predominantly in the gas phase. 
In the model without reverse shock destruction, the SFR increases from $\sim10^{8}\ {\rm yr}$ and saturates around $5\times10^{8}\ {\rm yr}$. 
This is because the gas is consumed by the star formation. 
In model A10m9 ($n_{\rm SN}=10\ {\rm cm^{-3}}$), SFR is suppressed until $t\sim0.8\times10^{9}\ {\rm yr}$. 

We should note that it is probably that $n_{\rm SN}<1\ {\rm cm}^{-3}$ for Pop III stars in the mass range of $20-40\ {\rm M}_{\odot}$, since Pop III stars are massive and can photoevaporate the clouds in which they form, and that ionized flows evacuate the dense gas around the stars to well below $n_{\rm SN}=1\ {\rm cm}^{-3}$ \citep{Wha04, Kit04}. 
In this case, considering circumstellar densities of $10\ {\rm cm}^{-3}$ and greater is not relevant to dust evolution in the SN remnant. 
In this paper, we consider $n_{\rm SN}>5\ {\rm cm}^{-3}$ for completeness. 
If the stars are forming at lower redshift and are enriched, they will have stellar winds that also sweep away circumstellar gas to low densities. 

We note that in usual star formation recipe in both numerical simulations and analytic models, SFR is assumed to increase with the local gas density and our results show that the SFR is strongly affected by $n_{\rm SN}$. 
We will discuss the effects of $n_{\rm SN}$ on the SFR in more detail in Section \ref{sec:sum}. 

In Figure \ref{fHtwoevo}, we show the change of the molecular fraction, $f_{{\rm H}_{2}}$, with metallicity, $Z$, for various $n_{\rm SN}$. 
In usual chemical evolution models, $Z$ is a key indicator of the galaxy evolution. 
However, as shown in this figure, $f_{{\rm H}_{2}}$ does not solely depend on the metallicity. 
For $Z\gtrsim5\times10^{-4}\ Z_{\odot}$, $f_{{\rm H}_{2}}$ is large in models with small $n_{\rm SN}$. 
This is because $D_{\rm area}/M_{\rm metal}$ is large (small) in models with small (large) $n_{\rm SN}$ for the same $Z$ as shown in Figure \ref{dustareatometal}. 

The molecular fraction is well described by the total dust cross-section per unit volume for $\sigma_{{\rm d},-20}\gtrsim0.001$. 
In Figure \ref{fHtwoevo2}, we show the evolution of molecular fraction in terms of total dust cross-section per unit volume. 
For $\sigma_{{\rm d},-20}\gtrsim1\times10^{-3}$, ${\rm H}_{2}$ formation on dust grains dominates $f_{{\rm H}_{2}}$ evolution as shown by the convergence of all models. 

\subsection{The dark matter halo mass}
\begin{figure}[!h]
\epsscale{1.0}
\plotone{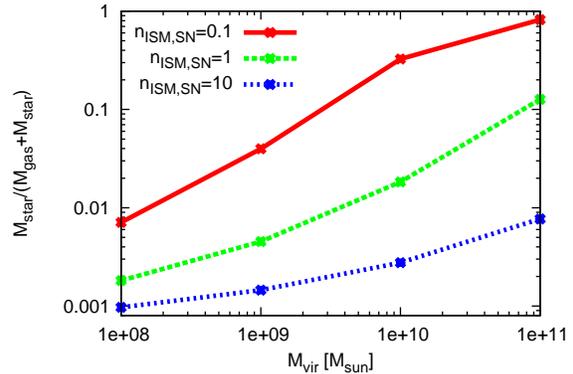}
\caption{The stellar mass fraction in the galaxy, $M_{\rm star}/(M_{\rm gas}+M_{\rm star})$, for the various virial masses, $M_{\rm vir}$ of the models with $n_{\rm SN}=0.1\ {\rm cm}^{-3}$ (solid line, model A0.1m8, A0.1m9 , A0.1m10, and A0.1m11), with $n_{\rm ISM.SN}=1.0\ {\rm cm}^{-3}$ (dashed line, model A1m8, A1m9, A1m10, and A1m11), and with $n_{\rm ISM.SN}=10\ {\rm cm}^{-3}$ (dashed line, model A10m8, A10m9, A10m10, and A10m10). }
\label{startogas}
\end{figure}
Finally, we show the stellar mass fraction, $M_{\rm star}/(M_{\rm gas}+M_{\rm star})$ for $M_{\rm vir}=10^{8}$, $10^{9}$, $10^{10}$ and $10^{11}$ $M_{\odot}$ at $z_{\rm vir}=10$, in Figure \ref{startogas}. 
We note that dark halos of virial masses, $M_{\rm vir}=10^{8}$, $10^{9}$, $10^{10}$ and $10^{11}$ $M_{\odot}$ correspond to the density fluctuation of $2.0\sigma$, $2.5\sigma$, $3.0\sigma$ and $4.1\sigma$, respectively. 
$M_{\rm star}/(M_{\rm gas}+M_{\rm star})$ is large for large $M_{\rm vir}$. 
This is explained as follows; the gas cools to the CMB temperature $\sim 30\ {\rm K}$ in all $M_{\rm vir}$ after $t\sim{\rm a\ few}\times10^{7}\ {\rm yr}$ although $T_{\rm vir}$ increases with $M_{\rm vir}$, so that the final gas density becomes higher because of smaller $H/r_{\rm disk}$ in larger $M_{\rm vir}$ (see Equation (\ref{eq:H})). 
This results in more rapid molecular formation in a larger $M_{\rm vir}$ halo (see Equation (\ref{fHtwoformation})). 
The rapid molecular formation enhances the star formation and as a results, causes the large stellar mass fraction. 
We should note that in higher $z_{\rm vir}$, $M_{\rm star}/(M_{\rm gas}+M_{\rm star})$ is larger for the galaxies with $T_{\rm vir}\gtrsim10^{4}\ {\rm K}$, since rotation timescale become smaller in our model. 

\section{Discussion and conclusions}
\label{sec:sum}
We have investigated the dust size evolution and the resulting ${\rm H}_{2}$ formation on dust grains in the galaxies in the early Universe ($z>5$). 
Our dust evolution model includes the dust production by SNe II and the effects of dust destruction by the reverse shocks and forward shocks driven by SNe. 
In the galaxy model, we follow the chemical network of ${\rm H}_{2}$ formation and the thermal evolution of gas taking into account cooling by ${\rm H}_{2}$, atomic hydrogen and metals, and heating by the stellar radiation. 
The SFR is assumed to be proportional to the mass of molecular hydrogen and the rotation timescale of galactic gas disk. 
The evolution of size distribution of dust has been taken into account for the first time in this paper to investigate the influence on star formation activity in galaxies. 

We obtain three main results. 
First, we show that ${\rm H}_{2}$ formation is suppressed by the dust destruction, especially by the reverse shocks in SNRs. 
Resulting molecular fraction in the galaxy model with dust destruction by both reverse shocks and forward shocks (see Figure \ref{0_5_1}) is 2.5 orders of magnitude less than the galaxy model without both destruction processes (see Figure \ref{0_6_2}) and is 2 orders less than the galaxy model considering dust destruction by only forward shocks (see Figure \ref{0_6_1}) for $n_{\rm SN}=1.0\ {\rm cm}^{-3}$. 
Our results show that dust size evolution has great effects on the early galaxy evolution. 

The dust destruction by reverse shocks is essential in our galaxy model and has more dominant role than forward shocks, since the reverse shock destroys small dust grains earlier than the forward shock. 
Dust destruction by a reverse shock becomes more (less) prominent for a higher (lower) $n_{\rm SN}$. 
In early galaxy evolution, considering dust destruction by reverse shocks is very important to ${\rm H}_{2}$ formation. 
We note that forward shocks affect dust size distribution in large dust-to-gas mass ratio, ${\cal D}\gtrsim10^{-3}$, as shown in Section \ref{res:dustdest}. 

Second, we show that the SFR strongly depends on the ISM density around SNe progenitors, $n_{\rm SN}$, which determines the efficiency of dust destruction by reverse shocks. 
As noted in Section \ref{sec:galmodel}, we treat $n_{\rm SN}$ different from the mean gas density in our one-zone galaxy model, $n_{\rm H}$, taking into account the effect of photo-ionization by SNe progenitors. 
The molecular fraction is different among the models with various $n_{\rm SN}$, even for the same $Z$ (see Figure \ref{fHtwoevo}), and it solely depends on total dust cross-section per volume from the time when ${\rm H}_{2}$ formation on dust grain becomes the most dominant process than the other processes of ${\rm H}_{2}$ formation and destruction (see Figure \ref{fHtwoevo2}). 
The ratio of total dust cross-section to total metal mass presented in Figure \ref{dustareatometal} is very useful for a numerical simulation of galaxy formation with dust size evolution. 

A higher ISM density around SN progenitors, $n_{\rm SN}$ results in lower SFR in the early galaxy evolution. 
In conventional galaxy formation theory, SFR has been assumed as an increasing function of gas density. 
Since in our one-zone model, we simply treat $n_{\rm SN}$ as a parameter, it is very interesting to resolve the ionized region around a SN progenitor by high-resolution radiation hydrodynamic simulation taking into account the effects of ionization heating by massive SNe progenitors. 

Finally, we show that galaxy evolution depends on $M_{\rm vir}$ and show that the stellar mass fraction, $M_{\rm star}/(M_{\rm gas}+M_{\rm star})$, is monotonically increasing functions of $M_{\rm vir}$. 
The halo number density in the redshift range $5<z<10$ is a decreasing function of the halo virial mass, $10^{8}\ M_{\odot}<M_{\rm vir}<10^{11}\ M_{\odot}$. 
In most papers on the reionization, $M_{\rm star}/(M_{\rm gas}+M_{\rm star})$ is assumed to be constant for various halo mass, $M_{\rm vir}$. 
The difference of stellar mass fraction with different $M_{\rm vir}$ in our results is important for galaxy evolution in the early Universe and should be considered in the cosmic reionization process. 
We will study this effects taking into account Population III star formation and the ${\rm H}_{2}$ dissociation by the Lyman-Werner background in a forthcoming paper, since strong Lyman-Werner background that is effective to dissociate molecular hydrogen, to delay gas condensation, and to suppress SF especially in the lower mass ${\rm H}_{2}$ cooling halos, is expected in the cosmic reionization era (Yamasawa et al. in preparation). 

Dust size distribution governs the extinction curve and reemitting IR spectrum \citep[e.g.][]{Sil98, Gra00, Tak05, Li08}. 
A ground-based interferometric facility, the Atacama Large Millimeter Array (ALMA)
\footnote{
http://www.almaobservatory.org/
} 
can be used to study high-redshift galaxies, since redshifted dust emission can be observed with ALMA. 
We can study dust emission and resulting SED of our galaxy model using our results of the dust size distribution and test our galaxy model by comparing to observational data in the future. 

Our galaxy model can be applied to the lower redshift $(z<5)$ star formation history, taking into account dust production by low-mass stars and growth of dust due to accretion of metal in ISM \citep{Ino03, Dra09}. 
The accretion timescale, $\tau_{\rm acc}$, depends on the metallicity, and is given by $\tau_{\rm acc}=\tau_{\rm acc,0}\ Z_{\odot}/Z$,
where $\tau_{\rm acc,0}\sim10^{8}\ {\rm yr}$ \citep{Ino03}. 
If we assume $Z\sim10^{-2}\ Z_{\odot}$ corresponding $\tau_{\rm acc}\sim10^{10}\ {\rm yr}$, then the accretion time is too long to affect dust size distribution for a cosmic time $\le1.2\ {\rm Gyr}$ ($z>5$). 
At a galaxy age of $\gtrsim1\ {\rm Gyr}$, dust production by low-mass stars should affect the dust size evolution, because after that epoch dust is supplied from low-mass stars as well as SNe II. 
We will include such processes to investigate low-redshift galaxies in the future work. 

Cosmological simulation of galaxy formation including our dust formation and evolution model is needed, 
since it is widely understood that most $z\sim10$ galaxies were not clear disk galaxies \citep{Joh08, Gre10, Wis10}. 
We will study first galaxy formation by cosmological simulation including our dust model. 

\acknowledgments
We thank the anonymous referee for very careful reading and very nice comments that improve this paper. 
The authors are grateful to M. Fujimoto, K. Sorai, K. Omukai, A. K. Inoue, T. Takeuchi, N. Yoshida and B. T. Draine for helpful discussions. 
Numerical computations were carried out on NEC SX-9 at the Center for Computational Astrophysics, CfCA, of National Astronomical Observatory of Japan. 
This work was partly supported by the Grant-in-Aid for Scientific Research of Japan Society for the Promotion of Sciences (08091823, 18104003, 0340038). 
H.H. is supported by NSC grant 99-2112-M-001-006-MY3. 
T.N. and K.N are supported by World Premier International Research Center Initiative, Next, Japan. 
\appendix


\begin{thebibliography}{}
\bibitem[Abel et al.(1997)]{Abe97} Abel, T., Anninos, P., Zhang, Y. \& Norman, M. L. 1997, \na, 2, 181
\bibitem[Bianchi \& Schneider(2007)]{Bia07} Bianchi, S. \& Schneider, R. 2007, \mnras, 378, 973 
\bibitem[Bigiel et al.(2008)]{Big08} Bigiel, F., Leroy, A., Walter, F., Brinks, E., Blok, W. J. G. D, Madore, B. \& Thornley, M. D. 2008, \apj, 136, 2846 
\bibitem[Barlow et al.(2010)]{Bar10} Barlow, M. J., Krause, O., Swinyard, B. M. and 13 authors 2010, \aap, 518, 138
\bibitem[Bromm et al.(2003)]{Bro03} Bromm, V., Yoshida, N. \& Hernquist, L. 2003, \apj, 596, 135 
\bibitem[Bromm et al.(2009)]{Bro09} Bromm, V., Yoshida, N., Hernquist, L. \& McKee, C. F. 2009, \na, 459, 49
\bibitem[Cazaux \& Spaans(2004)]{Caz04} Cozaux, S. \& Spaans 2004, \apj, 611, 40
\bibitem[Cherchneff \& Dwek(2010)]{Che10} Cherchneff, I. \& Dwek, E. 2010, \apj, 713, 1
\bibitem[Cole et al.(2000)]{Col00} Cole, S., Lacey, C. G., Baugh, C. M. \& Frenk, C. S. 2000, \mnras, 319, 168
\bibitem[Draine \& Lee(1984)]{Dra84} Draine, B. T. \& Lee, H. M. 1984, \apj, 285, 89
\bibitem[Cox(2000)]{Cox00} Cox, A. N. 2000, Allen's Astrophysical Quantities, 4th edn Springer, New York
\bibitem[Draine \& Bertoldi(1996)]{Dra96} Draine, B. T. \& Bertoldi, F. 1996, \apj, 468, 269
\bibitem[Draine(2009)]{Dra09} Draine B. T., 2009, in Henning Th., Gr\"un E., Steinacker J., eds, ASP Conf. Ser. Vol. 414, Cosmic Dust - Near and Far. Astron. Soc. Pac., San Francisco, p. 453 
\bibitem[Dwek et al.(2007)]{Dwe07} Dwek, E., Galliano, F., \& Jones, A. P. 2007, \apj, 662, 927
\bibitem[Dwek \& Cherchneff(2010)]{Dwe10} Dwek, E. \& Cherchneff, I. 2010, arXiv, 1011.1303v1
\bibitem[Ferrara et al.(2000)]{Fer00} Ferrara, A., Pettoni, M. \& Sjcjekinov, Y. 2000, \mnras, 319, 539
\bibitem[Ferrarotti \& Gail(2006)]{Fer06} Ferrarotti, A. S. \& Gail, H. -P. 2006, \aap, 447, 553
\bibitem[Gall et al.(2010)]{Gal10} Gall, C., Andersen, A. C. \& Hjorth, J 2010, arXiv, 1011.3157
\bibitem[Gallerani et al.(2010)]{Gal10} Gallerani, S., Maiolino, R., Juarez, Y. and 9 authors 2010, arXiv, 1006.4463
\bibitem[Galli \& Palla(1998)]{Gal98} Galli, D. G. \& Palla, F. 1998, \aap, 335, 403
\bibitem[Glover \& Abel(2008)]{Glo08} Glover, S. C. O. \& Abel, T. 2008, \mnras, 388, 1627 
\bibitem[Gnedin et al.(2009)]{Gne09} Gnedin, N. Y., Tassis, K. \& Kravtsov, A. V. 2009, \apj, 697, 55
\bibitem[Granato et al.(2000)]{Gra00} Granato, G. L., Lacey, C. G., Silva, L., Bressan, A., Baugh, C. M., Cole, S. \& Frenk, C. S. 2000, \apj, 542, 710
\bibitem[Greif \& Bromm(2006)]{Gre06} Greif, T. H. \& Bromm, V. 2006, \mnras, 373, 128 
\bibitem[Greif et al.(2010)]{Gre10} Greif, T. H., Glover, S. C. O., Bromm, V. \& Klessen, R. 2010, \apj, 716, 510 
\bibitem[Glover \& Jappsen(2007)]{Glo07} Glover, S. C. O. \& Jappsen, A. -K. 2007, \apj, 666, 1 
\bibitem[Haiman et al.(1996)]{Hai96} Haiman, Z., Thoul, A. A. \& Loeb, A. 1996, \apj, 464, 523
\bibitem[Herger \& Woosley(2002)]{Her02} Herger, A., \& Woosley, S. E. 2002, \apj, 567, 532
\bibitem[Herger et al.(2003)]{Her03} Heger, A., Fryer, C. L., Woosley, S. E., Langer, N. \& Hartmann, D. H. 2003, \apj, 591, 288
\bibitem[Hirashita \& Ferrara(2002)]{Hir02} Hirashita, H., \& Ferrara, A. 2002, \mnras, 337, 921
\bibitem[Hollenbach \& McKee(1979)]{Hol79} Hollenbach, D. \& McKee, C., F. 1979, \apj, 41, 555
\bibitem[Hollenbach \& McKee(1989)]{Hol89} Hollenbach, D. \& McKee, C. F. 1989, \apj, 342, 306
\bibitem[Hutchings et al.(2002)]{Hut02} Hutchings, R. M., Santoro, F., Thomas, P. A. \& Couchman, H. M. P. 2002, \mnras, 330, 927
\bibitem[Inoue (2003)]{Ino03} Inoue, A. K. 2003, \pasj, 55, 901
\bibitem[Jappsen et al.(2007)]{Jap07} Jappsen, A. -K., Glover, S. C. O., Klessen, R. S. \& Mac Low, M. -M. 2007, \apj, 660, 1332 
\bibitem[Jappsen et al.(2009a)]{Jap09a} Jappsen, A. -K., Mac Low, M. -M., Glover, S. C. O. \& Klessen, R. S. 2009, 694, 1161 
\bibitem[Jappsen et al.(2009b)]{Jap09b} Jappsen, A. -K., Klessen, R. S., Glover, S. C. O. \& Mac Low, M. -M. 2009, \apj, 696, 1065 
\bibitem[Joggerst et al.(2010a)]{Jog10a} Joggerst, C. C., Akmgren, A., Bell, J., Heger, A., Whalen, D. \& Woosley, S. E. 2010, \apj, 709, 11
\bibitem[Joggerst et al.(2010b)]{Jog10b} Joggerst, C. C., Akmgren, A. \& Woosley, S. E. 2010, \apj, 723, 353 
\bibitem[Johnson et al.(2008)]{Joh08} Johnson, J. L., Greif, T. H. \& Bromm, V. 2008, \mnras, 388, 26 
\bibitem[Jones et al.(1994)]{Jon94} Jones, A. P., Tielens, A. G. G. M., Hollenbach, D. J., \& McKee, C. F. 1994, \apj, 433, 797
\bibitem[Kennicutt(1998)]{Ken98} Kennicutt, R. C. 1998, \apj, 498, 541
\bibitem[Kitayama \& Ikeuchi(2000)]{Kit00} Kitayama, T. \& Ikeuchi, S. 2000, \apj, 529, 615
\bibitem[Kitayama et al.(2004)]{Kit04} Kitayama, T., Yoshida, N., Susa, H. \& Umemura, M. 2004, \apj, 613, 631
\bibitem[Kitayama \& Yoshida(2005)]{Kit05} Kitayama, T. \& Yoshida, N. 2005, \apj, 630, 675 
\bibitem[Kozasa et al.(2009)]{Koz09} Kozasa, T., Nozawa, T., Tominaga, N., Umeda, H., Maeda, K. \& Nomoto, K. 2009, ASP Conference Series, 414, 43
\bibitem[Kr\"ugel(2008)]{Kru08} Kr\"ugel, E. 2008, An Introduction to the Physics of Interstellar Dust, (Taylor \& Francis), ISBN 9781584887072 
\bibitem[Krumholz \& McKee(2005)]{Kru05} Krumholz, M. R. \& McKee, C. F. 2005, \apj, 630, 250
\bibitem[Li et al.(2008)]{Li08} Li, Y., Hopkins, P. F., Hernquist, L., Finkbeiner, D. P., Cox, T. J., Springel, V., Jiang, L. \& Yoshida, N. 2008, \apj, 678, 41
\bibitem[Maiolino et al.(2004)]{Mai04} Maiolino, R., Schneider, R., Oliva, E., Bianchi, S., Ferrara, A., Mannucci, F., Pedani, M. \& Roca Sogorb, M. 2004, \nat, 431, 533
\bibitem[Mackey et al.(2003)]{Mack03} Mackey, J., Bromm, V. \& Hernquist, L. 2003, \apj, 586, 1 
\bibitem[Machacek et al.(2001)]{Mac01} Machacek, M. E., Bryan, G. L. \& Abel, T. 2001, \apj, 548, 509 
\bibitem[Machacek et al.(2003)]{Mac03} Machacek, M. E., Bryan, G. L. \& Abel, T. 2003, \mnras, 338, 273 
\bibitem[Mathis et al.(1977)]{Mat77} Mathis, J. S., Rumpl, W. \& Nordsieck, K. H. 1977, \apj, 217, 425
\bibitem[Mo et al.(1998)]{Mo98} Mo, H. J., Mao, S. \& White, S. D. M 1998, \mnras, 295, 319
\bibitem[Nath et al.(2008)]{Nat08} Nath, B. B., Laskar, T. \& Shull, J. M. 2008, \apj, 682, 1055
\bibitem[Navarro et al.(1996)]{Nav96} Navarro, J. F., Frenk, C. S. \& White, S. D. M. 1996, \apj, 462, 563 
\bibitem[Nozawa et al.(2003)]{Noz03} Nozawa, T., Kozasa., Umeda, H., Maeda, K. \& Nomoto, K. 2003, \apj, 598, 785
\bibitem[Nozawa et al.(2006)]{Noz06} Nozawa, T., Kozasa, T. \& Habe, A. 2006, \apj, 648, 435
\bibitem[Nozawa et al.(2007)]{Noz07} Nozawa, T., Kozasa, T., Habe, A., Dwek, E., Umeda, H., Tominaga, N., Maeda, K. \& Nomoto, K. 2007, \apj, 666, 955
\bibitem[Nozawa et al.(2010)]{Noz10} Nozawa, T., Kozasa, T., Tominaga, N., Maeda, K., Umeda, H., Nomoto, K. \& Oliver, K. 2010, \apj, 713, 356 
\bibitem[Omukai(2000)]{Omu00} Omukai, K. 2000, \apj, 534, 809
\bibitem[Omukai et al.(2005)]{Omu05} Omukai, K., Tsuribe, T., Schneider, R. \& Ferrara, A. 2005, \apj, 626, 627
\bibitem[Omukai et al.(2010)]{Omu10} Omukai, K., Hosokawa, T. \& Yoshida, N. 2010, \apj, 722, 1793 
\bibitem[O'Shea \& Norman(2007)]{O'S07} O'shea, B. W. \& Norman, M. L. 2007, \apj, 654, 66 
\bibitem[O'Shea \& Norman(2008)]{O'S08} O'shea, B. W. \& Norman, M. L. 2008, \apj, 673, 14 
\bibitem[Pollack et al.(1994)]{Pol94} Pollack, L. B., Hollenbach, D., Beckwith, S., Simonelli, D. P., Roush, T. \& Fong, W. 1994, \apj, 421, 615
\bibitem[Robertson \& Kravtsov(2008)]{Rob08} Robertson, B. E. \& Kravtsov, A. V. 2008, \apj, 680, 1083
\bibitem[Salpeter(1955)]{Sal55} Salpeter, E. E. 1995, \apj, 121, 161
\bibitem[Schaerer(2002)]{Sch02} Schaerer, D. 2002, \aap, 382, 28
\bibitem[Schneider et al.(2004)]{Sch04} Schneider, R., Ferrara, A., \& Salvaterra, R., 2004, \mnras, 351, 1379
\bibitem[Schneider et al.(2006)]{Sch06} Schneider, R., Omukai, K., Inoue, A. \& Ferrara, A. 2006, \mnras, 369, 1437
\bibitem[Schneider \& Omukai(2010)]{Sch10} Schneider, R. \& Omukai, K. 2010, \mnras, 402, 429
\bibitem[Shakura \& Sunyaev(1988)]{Sha88} Shakura, N. I. \& Sunyaev, R. A. 1988, Adv.Sp.Res. 8, 135
\bibitem[Sibthorpe et al.(2010)]{Sib10} Sibthorpe, B., Ade, P. A. R., Bock, J. J., and 30 authors 2010, \apj, 719, 1553
\bibitem[Silva \& Denese(1998)]{Sil98} Silva, L. \& Danese, L. 1998, \apj, 509, 103
\bibitem[Silvia et al.(2010)]{Sil10} Silvia, D. W., Smith, B. D. \& Shull, J. M. 2010, \apj, 715, 1575
\bibitem[Smith et al.(2008)]{Smi08} Smith, B., Sigurdsson, S. \& Abel, T. 2008, \apj, 385, 1443 
\bibitem[Smith et al.(2009)]{Smi09} Smith, B. D., Turk, M. J., Sigurdsson, S., O'shea, B. W. \& Norman, M. L. 2009, \apj, 691, 441 
\bibitem[Spergel et al.(2007)]{Spe07} Spergel, D. N., Bean, R., Dor\'e, O. and 22 authors 2007, \apjs, 177, 377
\bibitem[Springel et al.(2005)]{Spr05} Springel, V, White, S. D. M., Jenkins, A., and 14 authors 2005, \na, 435, 629
\bibitem[Susa(2007)]{Sus07} Susa, H. 2007, \apj, 659, 908 
\bibitem[Susa(2008)]{Sus08} Susa, H. 2008, \apj, 684, 226 
\bibitem[Takeuchi et al.(2005)]{Tak05} Takeuchi, T., Ishii, T. T., Nozawa, T., Kozasa, T. \& Hirashita, H. 2005, \mnras, 362, 592
\bibitem[Tegmark et al.(1997)]{Teg97} Tegmark, M., Silk, J., Rees, M. J., Blanchard, A. \& Palla, F. 1997, \apj, 474, 1
\bibitem[Todini \& Ferrara(2001)]{Tod01} Todini, P. \& Ferrara, A. 2001, \mnras, 325, 726
\bibitem[Umeda \& Nomoto(2002)]{Ume02} Umeda, H. \& Nomoto, K. 2002, \apj, 565, 385
\bibitem[Valiante et al.(2009)]{Val09} Valiante, R., Schneider, R., Bianchi, S., \& Andersen, A. C. 2009, \mnras, 397, 1661
\bibitem[Wise \& Abel(2007)]{Wis07} Wise, J. H. \& Abel, T. 2007, \apj, 671, 1559 
\bibitem[Wise \& Abel(2008)]{Wis08} Wise, J. H. \& Abel, T. 2008, \apj, 685, 40 
\bibitem[Wise \& Cen(2009)]{Wis09} Wise, J. H. \& Cen, R. 2009, \apj, 693, 984
\bibitem[Wise et al.(2010)]{Wis10} Wise, J. H., Turk, M. J., Norman, M. L. \& Abel, T. 2010, arXiv, 1011.2632v2 
\bibitem[Yoshida et al.(2003)]{Yos03} Yoshida, N., Abel, T., Hernquist, L. \& Sugiyama, N. 2003, \apj, 592, 645 
\bibitem[Yoshida et al.(2008)]{Yos08} Yoshida, N., Omukai, K. \& Hernquist, L. 2008, Science, 321, 669 
\bibitem[Whalen et al.(2004)]{Wha04} Whalen, D., Abel, T. \& Norman, M. L. 2004, \apj, 610, 14 
\bibitem[Whalen et al.(2008)]{Wha08} Whalen, D., van Veelen, B., O'shea, B. W. \& Norman, M. L. 2008, \apj, 682, 49 
\bibitem[Wolfe \& Chen(2006)]{Wol06} Wolfe, A. M. \& Chen, H. -W. 2006, \apj, 652, 981 
\bibitem[Zhukovska et al.(2008)]{Zhu08} Zhukovska, S., Gail, H. -P. \& Trieloff, M. 2008, \aap, 479, 453
\end{thebibliography}
\end{document}